\documentclass[useAMS,usegraphicx]{mn2e}
\usepackage{rotate}
\usepackage{times}
\newif\ifAMStwofonts
\AMStwofontstrue

\def\gs{\mathrel{\hbox{\rlap{\hbox{\lower4pt\hbox{$\sim$}}}\hbox{$>$}}}}
\def\ls{\mathrel{\hbox{\rlap{\hbox{\lower4pt\hbox{$\sim$}}}\hbox{$<$}}}}

\def\chandra{{\it Chandra}}
\def\astroe{{\it Astro-E~2}}

\def\asca{{\it ASCA}}
\def\sax{{\it BeppoSAX}}
\def\xte{{\it RXTE}}
\def\xmm{{\it XMM-Newton}}
\def\ginga{{\it Ginga}}
\def\cgro{{\it CGRO}}

\def\et{{et al.\ }}
\def\mcg{{MCG--6-30-15}}
\def\3c{{3C~273}}

\def\rg{{\thinspace r_{\rm g}}}
\def\ka{{K$\alpha$}}
\def\ovii{{O~\textsc{vii}}}
\def\oviii{{O~\textsc{viii}}}
\def\fei{{Fe~\textsc{i}}}
\def\nh{{N_{\rm H}}}

\def\arcs{{\hbox{$^{\prime\prime}$}}}
\def\deg{^{\circ}}
\def\A{{\rm\thinspace \AA}}
\def\cm{{\rm\thinspace cm}}

\def\eV{{\rm\thinspace eV}}

\def\keV{{\rm\thinspace keV}}
\def\km{{\rm\thinspace km}}

\def\s{{\rm\thinspace s}}
\def\ks{{\rm\thinspace ks}}
\def\ps{{\rm\thinspace s^{-1}}}

\def\kmps{\hbox{$\km\ps\,$}}

\def\pscm{\hbox{$\cm^{-2}\,$}}

\title[A long, hard look at \mcg\ II]
      {A long, hard look at \mcg\ with \xmm\ II: detailed EPIC analysis
      and modelling}

\author[Vaughan \& Fabian]
       {S. Vaughan,$^{1,2}$ 
        A. C. Fabian$^{1}$ \\
$^{1}$ Institute of Astronomy, University of Cambridge, Madingley Road, Cambridge CB3 0HA\\
$^{2}$ X-Ray and Observational Astronomy Group, Department of Physics and Astronomy, University of Leicester, Leicester LE1 7RH
}
\date{Accepted 20/11/2003; submitted: 18/11/2003; in original form 21/9/2003}
\pagerange{\pageref{firstpage}--\pageref{lastpage}}
\pubyear{2003}
\begin{document}
\maketitle
\label{firstpage}

\begin{abstract}
The bright Seyfert 1 galaxy \mcg\ has provided some of the best
evidence to date for the existence of supermassive black holes in
active galactic nuclei.  Observations with \asca\ revealed an X-ray
iron line profile shaped by strong Doppler and gravitational effects.
In this paper the shape of the iron line, its variability
characteristics and the  robustness of this spectral interpretation
are examined using the long \xmm\ observation taken in 2001.  A
variety of spectral models, both including and excluding the effects
of strong gravity, are compared to the data in a uniform fashion. The
results strongly favour models in which the spectrum is shaped by
emission from a relativistic accretion disc.  It is far more difficult
to explain the $3-10 \keV$ spectrum using models dominated by
absorption (either by warm or partially covering cold matter),
emission line blends, curved continua or additional continuum
components.   These provide a substantially worse fit to the data and
fail to explain other observations (such as the simultaneous \sax\
spectrum).  This reaffirms the veracity of the relativistic `disc
line' interpretation.  The short term variability in the shape of the
energy spectrum is investigated and explained in terms of a
two-component emission model. Using a combination of spectral
variability analyses the spectrum is successfully decomposed into a
variable  power-law component (PLC) and a reflection dominated
component (RDC). The former is highly variable while the latter 
is approximately constant throughout the 
observation, leading to the well-known spectral variability patterns.
Consideration of the long term X-ray monitoring of
\mcg\ by \xte\ demonstrates that the long \xmm\ observation sampled
the `typical' state of the source.  These results and those of other
analyses of the large \xmm\ dataset are summarised and their
implications for understanding this enigmatic source are discussed.
\end{abstract}


\section{Introduction}
\label{sect:intro}

X-ray spectroscopy holds great promise for probing physics close to
the accreting black holes thought to power Active Galactic Nuclei
(AGN) and Galactic Black Hole Candidates (GBHCs). The high X-ray
luminosities observed from these  systems are thought to be generated
close to the black hole ($\ls 100 \rg$; where $\rg \equiv GM/c^2$ is the
gravitational radius) where the effects of strong gravity, such as
gravitational redshift and light bending, will imprint characteristic
signatures on the emerging X-ray spectrum.

As it escapes towards the observer, the primary X-ray emission will
interact with material in the innermost regions, modifying the
observed X-ray spectrum.  Reprocessing of the X-rays by an accretion
disc will lead to a particular set of spectral features. Absorption,
fluorescence, recombination, and Compton scattering in the surface
layers of the X-ray illuminated disc produce a `reflection'
spectrum. Under a wide range of physical conditions the most prominent
observables will be an emission line at
$\approx 6.4 \keV$ (the iron \ka\ fluorescence line) and a broad
`hump' in the spectrum which peaks in the range $20-40 \keV$
(e.g. Guilbert \& Rees 1988; Lightman \& White 1988; George \& Fabian
1991; Matt, Perola \& Piro 1991).  Observations of Seyfert 1 galaxies
with \ginga\ (Pounds \et 1990; Nandra \& Pounds 1994) demonstrated the
simultaneous presence of a $\approx 6.4 \keV$ emission line and an
up-turn in the spectrum above $\sim 10 \keV$, indicative of the
reflection hump.  The presence of the reflection hump has since been
confirmed by \cgro\ (e.g. Zdziarski \et 1995) and \sax\ (e.g. Perola
\et 2002).   

The bright Seyfert 1 galaxy \mcg\ ($z=0.007749$) has received
particular attention since \asca\ observations revealed the presence
of a broad, asymmetric emission feature at energies between $4 - 7
\keV$ identified with a highly broadened Fe \ka\ emission line (Tanaka
\et 1995).  Similar features were subsequently observed in other
Seyfert 1 galaxies (Mushotzky \et 1995; Nandra \et 1997).  The profile
of the line can be explained in terms of fluorescent emission from the
surface of  an accretion disc extending down to $\ls 6 \rg$ about a
supermassive black hole (SMBC): the relativistic `disc line' model
(Fabian \et 1989; Stella 1990; Laor 1991). For comparison, the
emission lines seen  in other wavebands (e.g. broad optical lines) are
thought to originate at distances $\gs 10^{3} \rg$. The broad Fe \ka\
line therefore potentially offers a powerful diagnostic of the
physical conditions in the immediate environment of the black hole
(see Fabian \et 2000; Nandra 2001 and Reynolds \& Nowak 2003  for
recent reviews).

Repeated observations of \mcg\ with \asca\ (Iwasawa \et 1996, 1999;
Shih \et 2002), \sax\ (Guainazzi \et 1999), \xte\ (Lee \et 1999, 2000;
Vaughan \& Edelson 2001), \chandra\ (Lee \et 2002) and \xmm\ (Wilms
\et 2001, hereafter W01, and Fabian \et 2002, hereafter paper~I) have confirmed the presence of
the broad, asymmetric emission feature. However, while this is
clearly not an artifact of any particular instrument or observation,
there are  uncertainties associated with the detailed modelling of the
spectrum.  The most challenging of these is that \mcg, along with many
other Seyfert 1s, shows complex absorption in its X-ray spectrum.
This complicates the process of identifying the correct underlying
continuum and thereby measuring the superposed line emission.  Indeed,
the presence of relativistically broadened  emission lines in the
X-ray spectra of other Seyfert 1 galaxies has recently become
something of a {\it cause celebre} (see e.g. Lubinski \& Zdziarski
2001; Inoue \& Matsumoto 2001; Branduardi-Raymont \et 2001; Pounds \&
Reeves 2002; Page, Davis \& Salvi 2003; Pounds \et 2003a,b).

The long ($\sim 320\ks$) \xmm\ observation of \mcg\ is  of great
importance to studies of the broad iron line, since  \mcg\ offers the
best established example of a relativistic line profile.  In
this paper the X-ray spectrum and spectral variability of \mcg\ are
examined using the long \xmm\ observation.  The interpretation of the
spectrum in terms of relativistic Fe \ka\ emission is  investigated by
fitting a range of models including both simple phenomenological
models and models accounting for the detailed physics expected in
photoionised accretion discs. These are compared to models that do not
include relativistic effects in order to test the strength of the
argument for strong gravity in \mcg.  A variety of techniques are
employed to characterise the spectral variability of the source and
thereby further constrain the range of  viable spectral models.  The
two \xmm\ observations are considered in the context of the long term
behaviour of \mcg\ using \xte\ monitoring observations spanning more
than six years.

The rest of this paper is organised as follows.
Section~\ref{sect:data} gives details of the \xmm\ and \xte\
observations and their data reduction.   Section~\ref{sect:history}
briefly reviews the \xte\ monitoring of \mcg\ and shows how the two
\xmm\ observations fit in with the long timescale properties of the
source.   Section~\ref{sect:fit} discusses the spectral fitting
results using models that include relativistic effects (Section
\ref{sect:fit_with}) and exclude relativistic effects (Section
\ref{sect:fit_without}).   In section~\ref{sect:spec_var} the spectral
variability of the source is examined using flux-flux plots
(section~\ref{sect:ff}), flux-resolved spectral analysis
(section~\ref{sect:fluxes}), rms spectra (section~\ref{sect:rms}),
time-resolved spectral analysis
(section~\ref{sect:time-res}) and Principal Component Analysis (PCA;
section~\ref{sect:pca}).
Section~\ref{sect:big_diff} discusses differences between the long
\xmm\ observation of 2001 and the previous observation taken in 2000.
Section~\ref{sect:review} summarises the new results from the present
work as well as the other analyses of  these \xmm\ data.  
Section~\ref{sect:disco} presents a discussion of the main results
and the main conclusions are given in section~\ref{sect:conc}.
Some details of the EPIC spectral calibration are
investigated in Appendix~\ref{app:cal}.


\section{Observations and data reduction}
\label{sect:data}

\begin{figure*}
\rotatebox{270}{ 
\scalebox{0.55}{\includegraphics{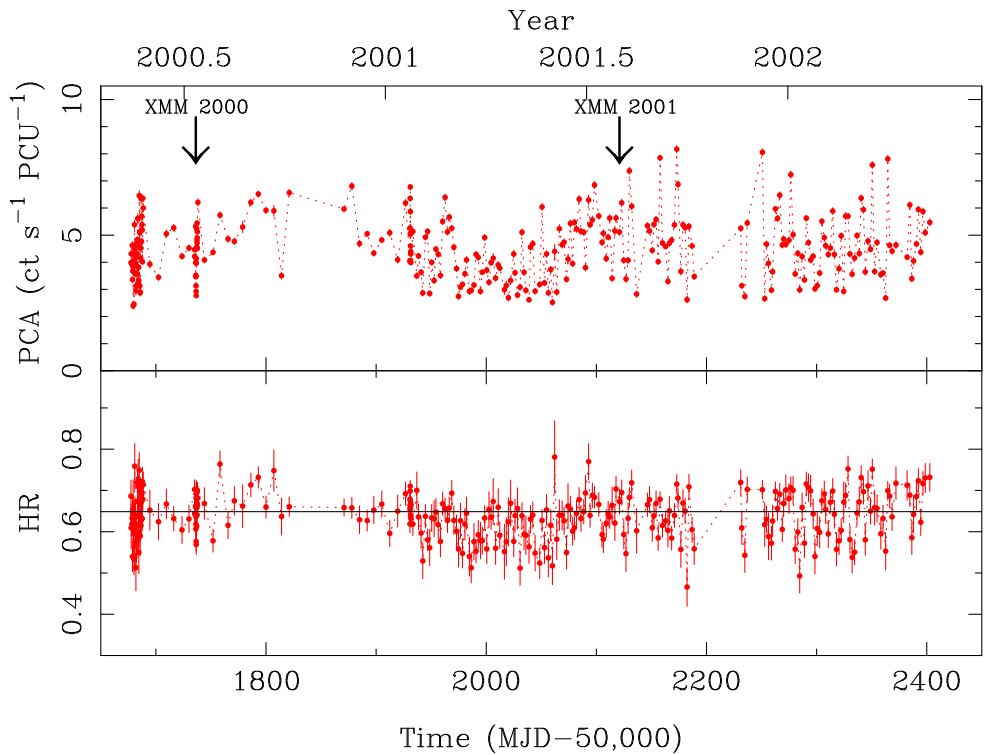}}}
\caption{ Top panel: the $2-10\keV$ \xte\ light curve  around the time
of the two \xmm\ observations (during PCA gain epoch 5).  Bottom
panel: Hardness ratio ($2-4/4-10\keV$). The solid line marks the
weighted mean hardness ratio throughout the observing period.  }
\label{fig:hr_big}
\end{figure*}

\subsection{\xmm\ observations}

\xmm\ (Jansen \et 2001) has observed \mcg\ twice. The first
observation was performed during revolution 108 (2000 July 11-12) and
the second observation was performed during revolutions 301--303 (2001
July 31 -- August 5).  In both cases the source was observed on-axis
and all instruments were operating nominally. The first results from
the 2000 \xmm\ observation were presented in W01 and the first results
from the 2001 \xmm\ observation were presented in paper~I.

During the 2000 observation the pn camera was  operated in small
window mode, the MOS2 camera in full frame mode and the MOS1 camera in
timing mode.  During the 2001 observation both the EPIC MOS cameras
and the EPIC pn camera were operated in small window mode.  For the
present analysis only the EPIC data taken in small window mode  were
used (i.e. only the pn from the 2000 observation, pn and both MOS from
the 2001 observation).  

The pn small window mode uses a $63 \times 64$
pixel window ($\approx 4^{\prime}.3 \times 4^{\prime}.4$ on the sky)
with the source 
positioned less than an arcminute from the CCD boundary.  The MOS
small window mode uses a window of $100 \times 100$ pixels  ($\approx
1'.8 \times 1'.8$) with the source image approximately centred  within
the window.  The small window modes allow CCD frames to be read out
every $5.7$~ms for the pn and $0.3\s$ for the MOS.  
These short frame times help lessen the effects of photon
pile-up which can distort the spectrum of bright sources (Ballet 1999;
Lumb 2000).   All three EPIC instruments used the medium filter.

\subsection{\xmm\ data reduction}

The extraction of science products from the Observation Data Files (ODFs)
followed standard procedures using the \xmm\ Science Analysis System
({\tt SAS v5.4.1}).  The EPIC data were processed using the standard
{\tt SAS}
processing chains to produce calibrated event lists.  These removed
events from the position of known defective pixels, corrected for
Charge Transfer Inefficiency (CTI) and applied a gain calibration to
produce a photon energy for each event.

Source data were extracted from circular regions of radius $40$\arcs
from the processed images and background events were
extracted from regions in the small window least effected by source
photons.  These showed the background to be relatively low and stable
throughout the observations, with the exception of the final few ks of
each revolution where the background rate increased (as the spacecraft
approached the radiation belts at perigee). Data from these
periods were ignored.   The total amount of `good' exposure time
selected was $64\ks$ from the pn during the 2000 observation, and for
the 2001 observation $315\ks$ and $228\ks$ from the MOS and pn,
respectively. (The 
lower pn exposure is due to the lower `live time' of the pn camera
in small-window mode, $\sim 71$ per cent; Str\"{u}der \et 2001).

The ratios of event patterns as a function of energy showed there is
negligible pile-up in the pn data but the MOS data suffer slightly
from pile-up. This may lead to a slight distortion of the MOS spectra.
In order to minimise this effect only photons corresponding to single
pixel events (EPIC patterns $0$)  were extracted from the MOS
cameras. The spectrum of single pixel events should be least affected
by pile-up (Molendi \& Sembay 2003). For the pn, single pixel (pattern
$0$) and double pixel (patterns $1-4$) events were extracted
separately for the spectral analysis and fitted simultaneously (the
spectrum of single pixel events provides better energy
resolution). Response matrices were generated using {\tt RMFGEN
v1.48.5} and ancillary responses were generated with {\tt ARFGEN
v.1.54.7}. 
For the 2001 observation source spectra were grouped such that each energy bin
contains at least $2,000$ counts for the pn and $400$ counts for the
MOS\footnote{This grouping was used instead of the more traditional
$N=20$ as it results in fewer spectral bins (and therefore reduces the
computation time for each fit iteration) but does not compromise the
intrinsic spectral resolution. The data contain sufficient counts that
even with this heavy binning the instrument resolution is
oversampled.}.  For the shorter 2000 observation the pn spectrum was
grouped to contain at least $500$ counts per energy bin.
Background subtracted light curves were extracted from the pn camera
in $1000\s$ bins using both single and double pixel events.

\begin{figure*}
\rotatebox{270}{
\scalebox{0.55}{\includegraphics{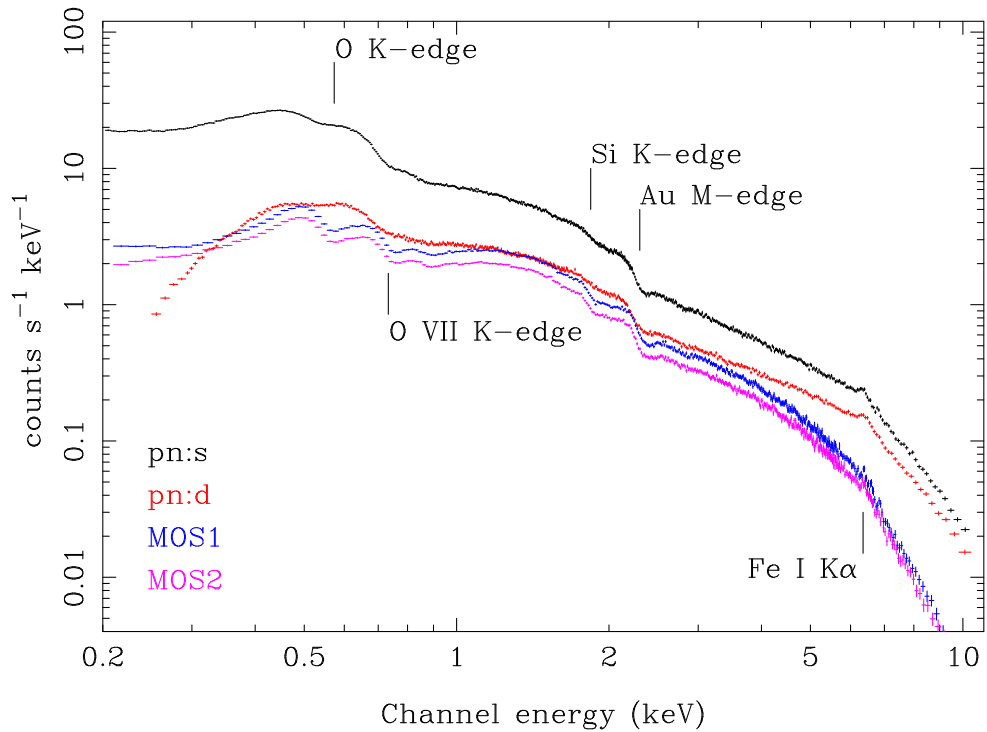}}}
\caption{
Full-band EPIC spectra. 
The four sets of data are the pn spectra for single and double
events (shown separately) and the MOS1 and MOS2 spectra.
For clarity the MOS2 spectrum has been shifted downwards by $20$ per cent.
Marked are the positions of the most significant instrumental
features (edges due to neutral O, Si and Au) and the positions of
the two features expected to be strong in the source spectrum
(the \ovii\ edge and Fe \ka\ emission line). 
}
\label{fig:spectrum}
\end{figure*}

\subsection{\xte\ observations}

\mcg\ has been regularly monitored by \xte\ since 1996.  The \xte\
Proportional Counter Array (PCA) consists of five collimated
Proportional Counter  Units (PCUs), sensitive  to X-rays in  a nominal
$2-60\keV$   bandpass   and   with   a   total   collecting   area   of
$\sim 6250\cm^{2}$. However, due to operational constraints there are
usually fewer than five PCUs operated during any given observation.
Only data from PCU2 were analysed here as this is the one unit that
has been operated consistently throughout the lifetime of the mission.
In the following analysis, data  from  the top  (most  sensitive)
layer  of  the  PCU array  were extracted  using {\tt  SAEXTRCT
v4.2d}.   Poor quality data were excluded using standard selection
criteria\footnote{ The acceptance criteria were as follows: {\sc
time\_since\_saa$>20$}~min; Earth   elevation  angle {\sc
elv$\geq 10\deg$}; offset from optical position of \mcg\ {\sc offset
$\leq 0.02\deg$}; and {\sc electron2 $\leq 0.1$}.  This  last
criterion removes data with high anti-coincidence  rate in the
propane layer of the  PCU.}.   The PCA background  was estimated
using  the latest  version of  the `combined' model   (Jahoda  \et
2000).  Light curves were  initially extracted from the {\tt
STANDARD-2} data with  16 s  time resolution.  The data were rebinned
into quasi-continuous observations lasting typically $1-2 \ks$.


\section{X-ray history of \mcg}
\label{sect:history}

Figure~\ref{fig:hr_big} shows the \xte\ light curve and hardness ratio
around the times of the two \xmm\ observations.
From this it is clear that the source displays strong X-ray
variability, but most is confined to short term `flickering' rather
than longer term trends in the data. A counter example can be seen
around MJD 52,000 when there appears to have been a prolonged period
of lower flux and variability, during which the spectrum
hardened slightly. However, both \xmm\ observations seem to have
caught \mcg\ in its `typical' state; there are no outstanding secular
changes in either flux or hardness ratio to suggest anything other
than normal behaviour in the source during the two \xmm\ observations.
The same conclusion is reached when the \xte\ monitoring extending
back to 1996 are considered.
Papadakis \et (2002) and Uttley, M$^{\rm c}$Hardy \& Papadakis (2002)
discuss the \xte\ monitoring observations in more detail.


\section{Spectral fitting analysis}
\label{sect:fit}

This section describes the results of the spectral fitting analysis.
The raw spectra from the 2001 observation are plotted in
Fig.~\ref{fig:spectrum}.  A variety of models, both including and
excluding relativistic effects, were fitted to the data. The goal of
the analysis was to find the models that can accurately match the
data, and test whether  models including relativistic effects were
preferred.  This analysis expands upon the analyses of the same
observation presented in paper~I, Fabian \& Vaughan (2003) and Ballantyne
\et (2003) in that the latest software and calibration were used (as
of 2003 May), a wider variety of spectral models was tested, and more
information from other observations was used to better constrain the
range of the model fits.


\subsection{Limits of spectral modelling}

The high signal-to-noise of these spectra mean that  systematic errors
in the detector response model can dominate over the statistical
errors (due to photon noise).  A brief analysis of some calibration
data is discussed in Appendix~\ref{app:cal}. This demonstrated that
significant instrumental features should be expected close to the
K-edges of O and Si and the M-edge of Au. Elsewhere the spectrum shows
no sharp 
instrumental residuals larger than $\sim 5$ per cent. Therefore in the
spectral fitting outlined below, the fitting is primarily carried out
over the $3-10\keV$ spectral range that excludes these features.
Appendix~\ref{app:cal} also demonstrates that there is a strong
discrepancy between the spectral slopes derived from each of the EPIC
cameras (see Molendi \& Sembay 2003). Therefore the spectral slope of
the power-law continuum was always allowed to vary independently
between the three EPIC cameras. 

With these precautions in place various spectral models were compared
to the data as described below.  The models were fitted to the four
spectra (pn:singles, pn:doubles, MOS1 and MOS2) simultaneously using
{\tt XSPEC v11.2} (Arnaud 1996).   The quoted errors on the derived
model parameters correspond to a $90$ per cent confidence level for
one interesting parameter (i.e. a $\Delta \chi^{2}=2.71$ criterion),
unless otherwise stated, and fit parameters are quoted for  the rest
frame of the source.  


\subsection{A first look at the spectrum}

\begin{figure}
\rotatebox{270}
{\scalebox{0.37}{\includegraphics{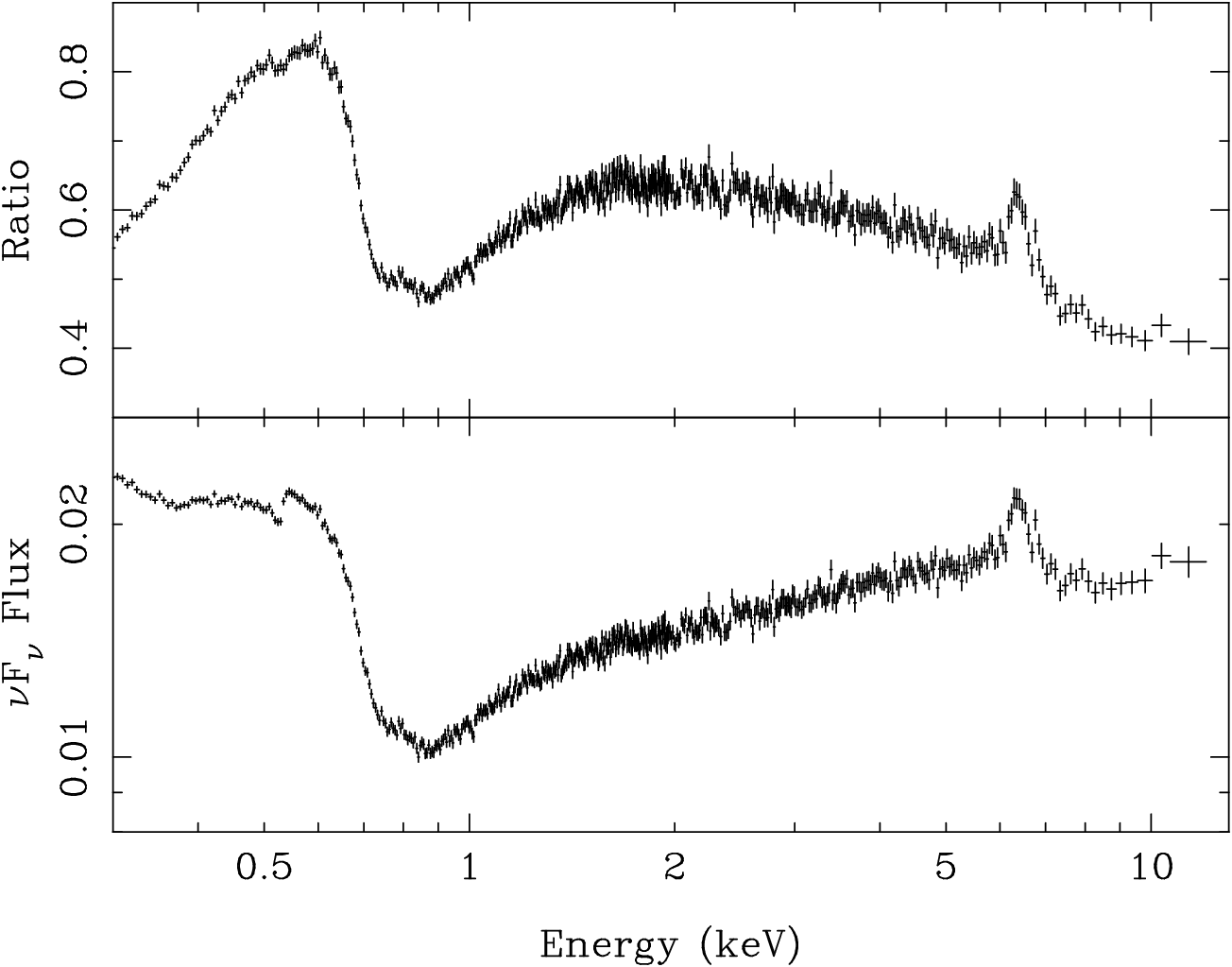}}}
\caption{
EPIC pn:s spectrum of \mcg\ after normalising by the \3c\ spectrum.
The top panel shows the ratio of counts in the \mcg\ and \3c\ spectra.
The bottom panel shows the ratio normalised by the spectral model
for \3c. This reveals the `fluxed' spectrum of \mcg\ (in units of ${\rm
keV}^{2} ~ {\rm s}^{-1} ~ {\rm cm}^{2} ~ {\rm keV}^{-1} $).
The `fluxed' spectrum has also been corrected for Galactic absorption.
(The sharp feature at $\approx 0.54\keV$ is an
artifact caused by the neutral O K-edge in the absorption
model being sharper than the instrumental response.) 
}
\label{fig:fluxed}
\end{figure}

The count spectra shown in Fig.~\ref{fig:spectrum} give only
a poor indication of the true spectrum of the source since they
are strongly distorted by the response of the detector. 
In order to gain some insight into the broad-band shape of the
source spectrum,
without recourse to model fitting, the EPIC data for \mcg\ were
compared to the spectrum of \3c\ (see Appendix~\ref{app:cal}).
The raw EPIC pn:s data (shown in Fig.~\ref{fig:spectrum}) were
divided by the raw pn:s spectrum for \3c. This gave the ratio
of the spectra for the two sources. 
\3c\ was chosen as it is a bright
source and has a relatively simple spectrum in
the EPIC band, i.e. a hard power-law plus smooth soft excess modified
by Galactic absorption (Page \et 2003b), and in particular does not
contain any strong, sharp spectral features such as lines or edges. 
Thus the ratio of \mcg\ to
\3c\ spectra will factor out the broad instrumental response to give 
a better impression of the true shape of the \mcg\ spectrum. The top
panel of Fig.~\ref{fig:fluxed} shows this ratio. The advantage of 
this plot is that it gives an impression of the \mcg\ spectrum and
is completely independent of spectral fitting. The disadvantages are 
that it is is in arbitrary units and only gives a very crude
representation of the true flux spectrum of \mcg\ because instead of
being distorted by the instrumental response, it is now distorted by
the spectrum of \3c. 

A more accurate, but
model-dependent method for obtaining a `fluxed' spectrum of \mcg\ is
to normalise the ratio by a spectral model for \3c.
This technique is very similar to the method routinely used in optical
spectroscopy of normalising the target spectrum using a standard star spectrum
to obtain a wavelength-dependent flux calibration.
The \mcg/\3c\ ratio was multiplied by a spectral model for \3c\
(defined in flux units). The spectral model comprised a hard power-law plus
two blackbodies to model the soft excess, modified by Galactic
absorption (see Appendix~\ref{app:cal} and also Page \et 2003b).
This process transformed the raw \mcg\ count spectrum into a `fluxed'
spectrum, without directly fitting the \mcg\ data. (The accuracy of
this procedure does however depend on the accuracy of the \3c\
spectral modelling.) The fluxed spectrum for \mcg\ is shown in the
bottom panel of Fig.~\ref{fig:fluxed}. 

Note that this procedure is not a formal deconvolution of the spectrum (see
Blissett \& Cruise 1979; Kahn \& Blissett 1980) and so does not 
correctly take into account the finite detector spectral resolution.
Nevertheless, it does provide a simple yet revealing impression of the
source spectrum. Specifically, the plot clearly reveals
the strongest spectral features in \mcg, namely the strong warm absorption
concentrated in the $\sim 0.7-2 \keV$ region and the iron
line peaking at $\approx 6.4 \keV$. 

\subsection{Ingredients for a realistic spectral model}

\subsubsection{Soft X-ray absorption}
\label{sect:wa_intro}

In the spectral fitting analysis described 
below the absorption due to neutral gas along the
line-of-sight through the Galaxy is accounted for  assuming a column
density of $\nh = 4.06 \times 10^{20}\pscm$ (derived from
$21\cm$ measurements; Elvis, Wilkes \& Lockman 1989).  Accurately
modelling the intrinsic absorption provided a more challenging
problem.  \mcg\ is known to possess a complex soft X-ray spectrum,
affected by absorption in partially ionised (`warm') material  (Nandra
\& Pounds 1992; Reynolds \et 1995; Otani \et 1996) and possibly also
dust (Reynolds \et 1997; Lee \et 2001; Turner \et 2003a; Ballantyne,
Weingartner \& Murray 2003).  However, the detailed structure of the
soft X-ray spectrum, as seen at improved spectral resolution with the
grating instruments on board \chandra\ and \xmm, remains controversial
(Branduardi-Raymont 2001; Lee \et 2001; Sako \et 2003; Turner \et
2003a).  In the present analysis the  complicating effects of the warm
absorber were mitigated by concentrating on the spectrum above
$3\keV$. However, this will not entirely eliminate the effects of
absorption.

In the $3-10 \keV$ range absorption by $Z\le 14$ ions will be limited
since their photoionisation threshold energies  are less than $3\keV$ for
all charge states (the ionisation energy for H-like Si$^{+13}$  is
$2.67\keV$; Verner \& Yakovlev 1995). Similarly, the L-shell
of ionised Fe  can significantly absorb the spectrum around $1\keV$
due to complexes of absorption lines and edges (e.g. Nicastro, Fiore
\& Matt 1999; Behar, Sako \& Kahn 2001). These transitions all occur below
$2.05 \keV$ (the ionisation energy of the L-edge in Li-like Fe$^{+23}$). 
Thus above $3\keV$ the  only opacity from low-$Z$ ions and
Fe~L will be due to the roll-over of their superposed photoelectric
absorption edges. The dominant effect  of the warm absorber on the
spectrum above $3\keV$ ($\lambda < 4.13\A$) is thus to impose some
subtle downwards curvature on the continuum towards lower energies.

The only significant absorption features occurring directly in the
$3-10\keV$ energy range will be due to K-shell transitions in the
abundant $Z>14$ elements (S, Ar, Ca, Fe and Ni). Absorption in the
$2-5.5\keV$ range could be due to S, Ar and Ca while Fe and Ni may
absorb above $6.40\keV$ and $7.47\keV$, respectively.  Comparison with
the high quality \chandra\  HETGS spectrum of the Seyfert 1 galaxy
NGC~3783, which shows a very strong warm absorber (Kaspi \et 2002),
suggests that absorption due to S, Ar and Ca is likely to be weak (see
the top panel of their figure 1 and also Blustin \et 2002, Behar \et
2003). This is confirmed by an analysis of the \chandra\ HETGS
spectrum of \mcg\ (J. Lee, piv. comm.; see also
section~\ref{sect:nh_test}).

Turner \et (2003a) fitted the RGS spectrum from the 2001 \xmm\
observation using a multi-zone warm absorber model plus absorption  by
neutral iron (presumably in the form of dust). Although the
transmission function of this absorption model recovers above a few
$\keV$ it nevertheless predicts the absorption remains significant at
higher energies (a $\sim 10$ per cent effect at $3\keV$) due to the
combined low-energy edges discussed above.  Therefore, to account for
the absorption in \mcg\ this multi-zone absorption model was included
in the  spectral fitting, with the column densities and ionisation
parameters kept fixed at the values derived from fits to the RGS data
(these are in any case very poorly constrained by the EPIC data above
$3\keV$).  The values used are tabulated in Table~1 of Turner \et
(2003a; model 2).  The photoelectric edges were included but not the
absorption lines.  This absorption model therefore accounts for the
subtle spectral curvature imposed by the warm absorber.   Resonance
absorption lines from S, Ar, Ca and Ni are unlikely to contribute
significant equivalent width.  The possibility of resonance line
absorption by Fe is explored separately below.

\subsubsection{Iron K-shell absorption}
\label{sect:fe_abs}
 
The other noteworthy effects of the warm absorbing material above
$3\keV$ are due to the K-shell of iron.  Iron in F-like to H-like ions
(${\rm Fe}^{+17}-{\rm Fe}^{+25}$) can absorb through \ka\ resonance
lines  in the $6.4 - 6.9\keV$ range (Matt 1994; Matt, Fabian \&
Reynolds 1997) which would be poorly resolved (if at all) in the EPIC
spectra. Sako \et (2003) estimated the resonance lines could produce a
total absorption equivalent width of $EW \sim -30\eV$ in \mcg\ based
on an analysis of the Fe L-shell transitions present in the RGS
spectrum from the 2000 \xmm\ observation.  This absorption could alter
the shape  of the observed emission around the Fe \ka\ emission line
and was therefore accounted for in the analysis (paper~I briefly
explored the possible effect of iron resonance absorption).

The other major source of opacity is K-shell photoelectric absorption  of
photons above $7.1\keV$ (Palmeri \et 2002).  This could affect the
continuum estimation on the high energy side of the line, and
therefore the apparent line profile, if not properly accounted for
(see discussion in e.g. Pounds \& Reeves 2002). The edge due to
neutral iron is negligible in \mcg;  the Galactic and intrinsic (dust)
neutral absorption gave $\tau \ls 0.01$.  The warm absorbing material
contains Fe in wide a range of charge states, leading to a blurring
together of the many K-edges over the $\sim 7.1 - 9\keV$ range (see
Palmeri \et 2002). In the Turner \et (2003a) absorption model the total
optical depth due to these edges (estimated by comparing the model
transmission below the \fei\ edge to that at the minimum of the
resulting absorption trough) is only $\tau_{\rm max} \sim 0.02$. Using
the ionic columns for Fe derived by Sako \et (2003) gave similar
results ($\tau_{\rm max} \sim 0.01$ around the K-edges).

This suggests that while the
line-of-sight material does  produce iron K-edge absorption, it is
rather shallow and spread over the $7-9\keV$ range. 
The resulting distortion on the transmitted spectrum is far
smaller than the statistical errors on the data above $7\keV$ and thus
is of little consequence to the spectral fitting. For completeness the
absorption due to dust and warm gas in \mcg\ (as derived by Turner \et
2003a) was included in the absorption model used below.  Furthermore,
the $7-10\keV$ EPIC data were searched for additional Fe absorption
(see section~\ref{sect:fe_test}), such as Fe~{\sc xxv - xxvi} edges
which would be undetectable in the lower energy grating spectra (due
to the lack of L-shell electrons).

\subsubsection{Reflection continuum}
\label{sect:ref_intro}

Observations of \mcg\ extending beyond $10\keV$ with \ginga\ (Pounds \et
1990; Nandra \& Pounds 1994), \xte\ (Lee \et 1998, 1999) and \sax\
(Guainazzi \et 1999; paper~I) revealed an upturn in the spectrum and
indicated the presence of a reflection continuum.  Particularly
relevant to the present analysis is that \sax\ observed \mcg\
simultaneously with the 2001 \xmm\ observation. The data from the PDS
instrument provided spectral information extending up to $\sim
100\keV$. Paper~I and Ballantyne \et (2003) presented the results of
fitting the \sax\ data with reflection models (see in particular
section 3 of Ballantyne \et 2003). These results strongly indicated
that the reflection continuum is strong ($R \gs 2$). 

\subsubsection{Narrow iron emission line}
\label{sect:core_intro}

As discussed in Section~\ref{sect:intro}, various missions have
resolved the broad emission feature spanning the $4 - 7\keV$ range.
In each observation the spectrum can be fitted in terms of a disc
line. There could in principle also be a contribution to the
reflection spectrum from more distant material ($\gg 100~\rg$) which
would produce a much narrower, symmetric line. Such lines have been
resolved in the spectra of many other Seyfert 1 galaxies (e.g. Yaqoob
\et 2001; Yaqoob, George \& Turner 2002; Pounds \et 2003a; Reeves \et
2003). The presence of such a narrow `core' could significantly affect
model fits to the iron line profile if not correctly accounted for
(e.g. Weaver \& Reynolds 1998).

In the case of \mcg\ the variations in the line profile observed by
\asca\ during a low-flux period (the `deep minimum') suggested the
contribution to the line from distant material was small (Iwasawa \et
1996). The higher resolution \chandra\ HETGS spectrum (Lee \et 2002) was
able to resolve the peak of the line emission near $6.4\keV$ and
confirm that the contribution due to distant material (i.e. an
unresolved line `core') is weak ($EW \ls 20\eV$).  The 2000 \xmm\
observation of \mcg\ also showed evidence for narrow line emission
(W01; Reynolds \et 2003).  Therefore the possibility of an unresolved
iron emission line is  allowed in the spectral models considered below.


\subsection{Fitting the Fe line core}
\label{sect:line_core}

\begin{figure}
\rotatebox{0}
{\scalebox{0.475}{\includegraphics{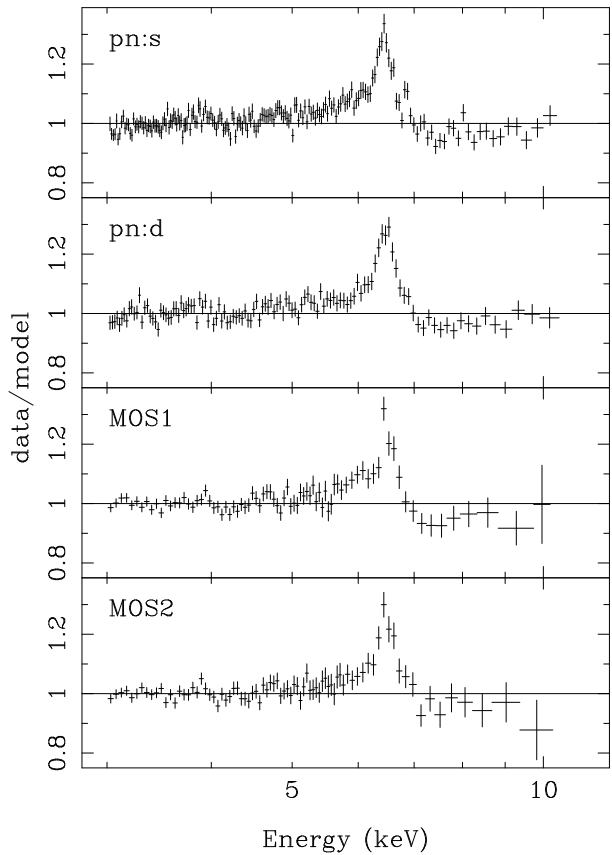}}}
\caption{
Ratios of EPIC spectra to absorbed power-law model. 
The model was fitted to the $3-5$ and $8-10\keV$ data.
Note the model underestimates the flux on the
red side of the line ($\sim 5-6.4\keV$) while the flux is 
overestimated above $7\keV$. Also note the narrow `notch'
at $\approx 6.7\keV$ noticeable in the pn:s spectrum.
}
\label{fig:spectrum_rat}
\end{figure}

Figure~\ref{fig:spectrum_rat} shows residuals when the EPIC  spectra
are fitted with a simple model comprising a power-law plus neutral and
warm absorption (as discussed in section~\ref{sect:wa_intro}).  The
only free parameters were the power-law slope (photon index $\Gamma$)
and normalisation.  The spectral interval immediately around the Fe K
features ($5-8\keV$) was ignored during the fitting, and the model was
subsequently interpolated over these energies.  The residuals clearly
reveal the strong iron \ka\ emission line peaking near $6.4\keV$.
The emission line appears asymmetric in this ratio plot,
extending down to at least $\sim 5\keV$ (this red wing on the line is
far broader and stronger than that expected from the Compton shoulder
from cold gas;
Matt 2002).

As a first step towards modelling the iron features the $5-8\keV$
region was examined in detail.   A model was built in terms of a
power-law  continuum (absorbed as described in
section~\ref{sect:wa_intro}) plus three Gaussian lines.  The
best-fitting parameters are shown in Table~\ref{table:core}.  The
lines were added to model the resolved line emission (line 1), the
unresolved absorption at $\approx 6.7\keV$ (line 2) and the unresolved
$6.4\keV$ emission line (line 3). Modelling the line with the single
resolved Gaussian emission line provided a reasonable fit, but the
inclusion of the unresolved absorption and emission lines
significantly improved the quality of the fit.  The inclusion of the
absorption line was an attempt to account for possible Fe resonance
absorption lines (section~\ref{sect:fe_abs}).

\begin{table}
\caption{
Best-fitting parameters for Gaussian models
of the iron line core region ($5 - 8 \keV$).
$^{f}$ indicates the parameter was fixed.
}
\centering
\label{table:core}            
\begin{tabular}{@{}l|rrrr@{}}
\hline
Line & $E(\keV)$ & $\sigma (\eV)$ & $EW (\eV)$ & $\chi^{2}/dof$ \\
\hline
1        &  $ 6.38\pm0.02$ & $291\pm33$ & $145\pm15$ & $ 318.5/291$   \\
\hline
1        & $6.41\pm0.03$ & $312\pm33$ & $167\pm20$  &   $300.2/289$        \\
2        & $6.73_{-0.04}^{+0.01}$ & $10^{f}$ & $-11\pm6$ &         \\
\hline
1        & $6.35\pm0.05$ & $405\pm50$ & $148\pm20$ &  $ 263.3/288$ \\
2        & $6.74_{-0.09}^{+0.15}$ &   $10^{f}$ &  $\ge-8.6$ &   \\
3        & $6.40^{f}$ & $10^{f}$      & $20\pm5$   &        \\
\hline
\end{tabular}
\end{table}

\begin{figure}
\rotatebox{270}
{\scalebox{0.32}{\includegraphics{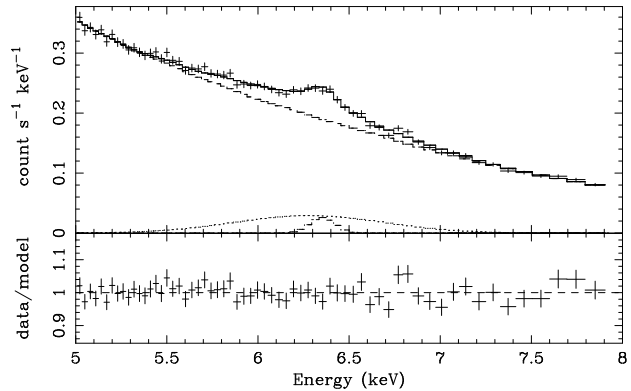}}}
\caption{
EPIC pn:s spectrum immediately around the
core of the Fe line (crosses). 
The histograms show the spectral model components, namely 
a power-law continuum, a broad
($\sigma \approx 410\eV$) Gaussian emission
line at $\approx 6.4\keV$, an unresolved emission line at $6.4\keV$ and
an unresolved absorption feature at $\approx 6.7\keV$.
}
\label{fig:line_core}
\end{figure}

Figure~\ref{fig:line_core} shows the spectrum around the peak of the
Fe emission fitted with the three line model (only EPIC pn:s data are
shown but  all four spectra were included in the fitting).  The
resolved emission line accounted for the majority of the flux from the
core of the Fe emission line, and was significantly resolved with a
velocity width $FWHM \approx 4.5\times10^4$~\kmps. Fitting the four
EPIC spectra individually with the same model gave consistent results;
in each case the emission line was very significantly resolved and the
measured widths were consistent with one another.

The high dispersion velocity width of the emission line core is robust
to the details of the underlying continuum fitted over this fairly
narrow energy range (it remain significantly resolved after including
additional cold gas and/or Fe~K edge absorption in the model; see also
section~\ref{sect:fe_test}).  Forcing the entire emission line to be
narrow (i.e. unresolved) gave an unacceptable fit ($\Delta \chi^2 =
+266.2$ compared with the resolved line).   In addition to $\approx
6.4\keV$ emission there could also be Fe lines at energies $\approx
6.7$ and $\approx 6.9\keV$, corresponding He-like and H-like ions,
respectively.  The simultaneous presence of iron lines at these three
energies (convolved through the EPIC response) could in principle
conspire to make the Fe emission mimic a single, resolved line.  See
Matt \et (2001) and Bianchi \et (2003) for an examples where such line
blends have been observed in EPIC spectra of Seyferts.  Allowing for
the presence of three unresolved lines (at energies of  $\approx 6.4$,
$\approx 6.7$ and $\approx 6.9\keV$) also gave an unacceptable
fit. The Fe emission was thus unambiguously resolved by EPIC.

This is of great interest because the resolved line core is
considerably broader than those of some other Seyfert 1  galaxies
(e.g. Yaqoob \et 2001; Kaspi \et 2002; Pounds \& Reeves 2002; Page \et
2003a; Page \et 2003c)  and suggests an origin close to the central
SMBH. 
Neglecting for the time being the contribution of any strongly redshifted
component to the  line profile, the width of the resolved core can be used to
infer the radial distance of the line emitting gas under the
assumption that the material is gravitationally bound (and emits at
$6.40\keV$ in the rest frame). In this case the rms velocity width of
the line ($\sigma$) is determined by its proximity to the SMBH
($r/\rg$) and the geometry according to: $r/\rg \sim
(c/\sigma)^2 /q$ (where $q$ depends on the geometry; Krolik
2001). Assuming randomly oriented circular orbits ($q=3$) places the line
emitting material at $r \sim 80\rg$, while assuming a Keplerian
disc (inclined at $i=30\deg$; $q=8$) requires the material to be at
$r \sim 30\rg$. The resolved line core thus suggests emission from
near the central SMBH, necessitating the use of models accounting
for the relativistic effects operating in this region.


Fitting this model over the $3-10\keV$ range gave a bad fit
($\chi^{2} = 983.1/ 777~dof$) with the emission line energy fixed to
$6.4\keV$. Allowing the energy of the broad Gaussian to be free
improved the fit ($\chi^{2}= 862.1/776~dof$) but the Gaussian became
extremely broad and redshifted ($E=5.9\pm0.2\keV$ and $\sigma \gs
0.7\keV$). Fitting with two Gaussians to model the broad line (to
account for the resolved core and also a highly redshifted component), 
as well as the narrow $6.4\keV$ emission and $\approx 6.7\keV$
absorption, provided a good fit ($\chi^{2}=730.7/773~dof$).
The two broad Gaussians had the following parameters:
$E_1=6.38_{-0.04}^{+0.06}\keV$, $\sigma_1=352_{-50}^{+106}\eV$,
$EW_1\approx 120\eV$ and $E_2=4.9_{-1.5}^{+0.2}\keV$,
$\sigma_2=1.0_{-0.2}^{+0.4}\keV$, $EW_2\approx 150\eV$.  
The energy and width of the second Gaussian further suggests the
presence of a highly broadened and 
redshifted component to the Fe emission. However, such
a claim is dependent on the assumed form of the underlying continuum.


\subsection{Dependence on the continuum model}
\label{sect:simple}

The simple absorbed power-law continuum model used above
(Fig.~\ref{fig:spectrum_rat}) 
clearly overestimated the continuum above $7\keV$ (on the `blue' side
of the line). These residuals are a direct result of fitting the
continuum with an over-simplistic model, and specifically could be
caused by of one of the following:
($a$) additional K-shell absorption by Fe, 
($b$) intrinsic curvature in the continuum, 
($c$) a continuum curved by excess absorption, or 
($d$) excess emission on the red side of the line ($3-5.5\keV$).
These last two possibilities would
lead to the true continuum slope over the $3-10\keV$
range being underestimated (the fit was driven by the better
statistics in the $3-5\keV$ region). 

\begin{figure}
\rotatebox{270}
{\scalebox{0.35}{\includegraphics{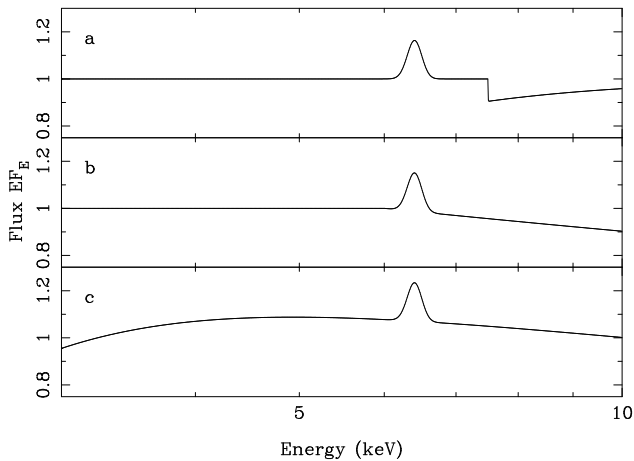}}}
\caption{
Schematic of three alternative models that do not include an emission
feature on the red side of the iron line.  The models comprise a
power-law continuum with resolved Gaussian line in addition to: ($a$)
additional Fe K-edge absorption, ($b$) a break in the power-law slope
at $\approx 6\keV$ and ($c$) enhanced soft X-ray absorption.
}
\label{fig:models}
\end{figure}

The first three of these alternative possibilities ($a-c$), illustrated
in figure~\ref{fig:models}, are briefly
discussed below and found to be unsatisfactory.  In the subsequent
analysis of sections~\ref{sect:fit_with} and \ref{sect:fit_without} it
was assumed that there is excess emission extending from the Fe line
to lower energies (i.e. possibility $d$), and this was modelled using
relativistic and non-relativistic emission features, respectively.

\subsubsection{Additional iron K absorption?}
\label{sect:fe_test}

As discussed in section~\ref{sect:fe_abs} the  Fe K-shell absorption
edges expected based on the grating spectra has been included in the
absorption model. It remains possible that there is absorption from
additional Fe that was not accounted for in this model.

\begin{figure}
\rotatebox{270}
{\scalebox{0.32}{\includegraphics{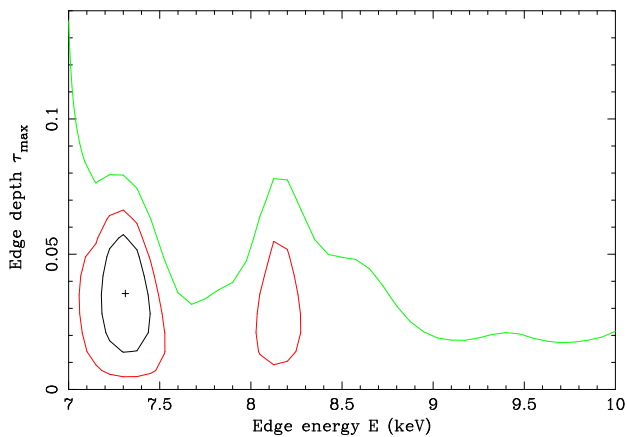}}}
\caption{
Confidence contours ($68.3, 90, 99$ per cent limits)
on absorption edge parameters when the $7-10\keV$
spectrum was searched for absorption.
}
\label{fig:Fe_edge}
\end{figure}

The $7-10\keV$ EPIC spectrum was searched for additional  absorption
by fitting the data over only this limited energy range with a
power-law and an absorption edge.  Although this model is very simple,
over the limited energy range the underlying continuum should not
deviate noticeably from a power-law, and with the limited resolution
and statistics available at these energies an edge can provide an
approximate description of additional Fe absorption.

The data constrain the depth of any such edge to be $\tau < 0.06$
($90$ per cent confidence) over the energies expected for Fe K
($7.11-9.28\keV$). In particular the depth of edge allowed at the
energies expected for Fe~{\sc xxv - xxvi} (at the rest-frame velocity
of \mcg) is $\tau \ls 0.03$, ruling out the presence of a large column
of such gas (which would have gone undetected in the low energy
grating spectra).

The best fitting edge model gave $E=7.31\pm0.14\keV$
and $\tau=0.036\pm0.024$.  Figure~\ref{fig:Fe_edge} shows the
confidence contours for the edge parameters in this `sliding edge'
search.  Including this edge in the model made virtually no difference
to the residuals around the iron line. In particular the apparent
asymmetry remained. That said, the best-fitting optical depth should
perhaps be treated as an upper limit due to the likely presence of Fe
K$\beta$ emission at $E \approx 7.06\keV$, which must accompany the
\ka\ emission. Allowing for the possibility of line emission at
$E = 7.06\keV$ resulted in the  edge depth becoming consistent with
zero, with a limit of $\tau \le 0.06$.  Including the Fe K edge absorption
contained in the model discussed  in section~\ref{sect:wa_intro}
further reduced the depth of any excess Fe absorption.

A caveat is that if a large column of Fe is present with a range of
intermediate charge states the resulting absorption  will not resemble
an edge (Palmeri \et 2002). However, the low energy grating data do
not support the presence of such a large column of ionised Fe in addition
to that already included in the model (Lee \et 2001; Sako \et 2003;
Turner \et 2003a).  

\subsubsection{An intrinsically curved $3-10\keV$ continuum?}
\label{sect:bkn_test}

Ballantyne \et (2003) and paper~I presented analyses of the simultaneous \sax\ PDS
spectrum of \mcg. After allowing for the
reflection continuum, the underlying power-law did not show any
evidence for intrinsic curvature. In particular, the lower limit on
the energy of an exponential cut-off in the continuum was constrained
to be $E_{\rm cut} \gs 100\keV$. Thus any subtle spectral curvature due
to the continuum deviating from a power-law at high energies (as
predicted by thermal Comptonisation models) will not affect the
$3-10\keV$ bandpass.

As a check for the possible effect of more drastic curvature in the
$3-10\keV$ band, the EPIC spectrum was fitted with a broken power-law
model, after excluding the region immediately around the Fe K features
($5-8\keV$). This provided a good fit ($\chi_{\nu}^{2} = 0.92$). The
break in the continuum occurred at $E=4.85_{-0.20}^{+0.86}\keV$ where
the slope changed from $\Gamma_{1}=2.00_{-0.04}^{+0.02}$ to
$\Gamma_{2}=2.14_{-0.03}^{+0.06}$ above the break energy (the slopes
are given based on the pn:s spectrum, similar differences were derived
from the other spectra). This form of spectral break, with the
spectral slope increasing by $\Delta \Gamma \approx 0.15$ above
$5\keV$ is highly unusual and unexpected when compared to other
Seyfert 1 spectra.  Furthermore, the strong reflection component  seen
in the \sax\ PDS data (section~\ref{sect:ref_intro}) should lead to a
flattening of  the spectral slope at higher energies. Indeed, when the
broken power-law model was  interpolated across the $5-8\keV$ region
it predicted a broad iron line extending down to at least $5\keV$ (the
energy of the continuum break) but appeared to match poorly the
continuum slope at energies above the line (the model was too steep
compared to the data above $7\keV$).  An intrinsically curved/broken
$3-10\keV$ continuum is therefore considered highly unlikely.

\subsubsection{Excess soft X-ray absorption?}
\label{sect:nh_test}

\begin{figure}
\rotatebox{0}
{\scalebox{0.475}{\includegraphics{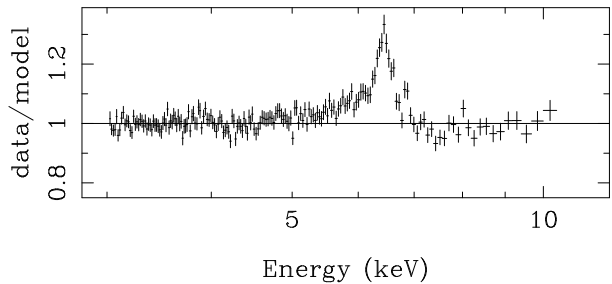}}}
\caption{
Ratio of EPIC pn:s spectrum to an absorbed power-law model.  The model
was fitted over the ranges $3-5\keV$ and $8-10\keV$  with the
absorption model discussed in section~\ref{sect:wa_intro}  but
allowing the parameters of one of the absorbing zones to be free (see
section~\ref{sect:nh_test}).  There is little difference in the
pattern of residuals compared with figure~\ref{fig:spectrum_rat}.  In
particular there is excess emission on the red side of the line
extending down to at least $5\keV$, and again the model slightly
overestimates the flux above $7\keV$.
}
\label{fig:spectrum2_rat}
\end{figure}

The warm absorbing gas could perhaps impose curvature on the $3-5.5\keV$
continuum in excess of that already accounted for in the absorption model
(section~\ref{sect:wa_intro}).  
Significantly higher columns of O or Si ions, for example, would
have the effect of increasing the opacity in the high energy tail of
the absorption edges occurring at energies $\ls 3\keV$.  This can be
quite effectively modelled by simply allowing the parameters of one of
the warm absorbing zones (namely the column density and ionisation
parameter) to be free in the fitting. The data were therefore fitted
in the $3-10\keV$ range, again excluding Fe K region ($5-8\keV$),
using a power-law modified by the absorption model of
section~\ref{sect:wa_intro} but allowing the parameters of one of the
warm absorber zones to vary.  The result was that column density of
the low ionisation absorber increased to fit better the slight
curvature in the $3-5\keV$ band.  The fit was good ($\chi_{\nu}^{2} =
0.92$) but when the model was interpolated into the Fe K region the
residuals again suggested a strongly asymmetric emission feature (as
shown in Figure~\ref{fig:spectrum2_rat}).  
The inclusion of extra absorption below $3\keV$ did therefore
not alter the requirement for an asymmetric, red emission
feature. However, this model is disfavoured as it severely
under-predicted the flux level at lower energies when extrapolated
below $3\keV$. Very similar results were obtained after allowing the
other (higher ionisation) absorbing zones to vary during the fitting. 

The other possibility, namely that the $3-5.5\keV$ spectrum is
significantly affected by S, Ar and/or Ca absorption,
was also explored. Absorption by the He- and H-like states of these ions
will not affect the spectrum $<2\keV$ (except if they are accompanied
by significant columns of M- and/or L-shell ions; Behar \& Netzer
2002) and therefore the soft X-ray EPIC and RGS data do not strongly constrain
their presence. The upper limit on the optical depth of an absorption
edge is $\tau_{\rm max} \ls 0.02$ throughout most of this energy
range. The only exception is at $\approx 4\keV$ where the addition of
a weak edge ($\tau \approx 0.03$) did significantly improve the
fit. Thus it remains plausible that some absorption by, for example,
highly ionised Argon (e.g. Ar~{\sc xvii}) may weakly contribute in
this region. However, this feature must be considered with caution. 
An examination of the EPIC calibration data for \3c\ (see
Appendix~\ref{app:cal}) revealed similar residuals at the same (observed
frame) energy. This feature could thus be due to an instrumental
effect at the $\approx 3$ per cent level (possibly due to 
Ca residuals on the mirror surfaces; F. Haberl priv. comm.).
In any event the affect on the derived iron line parameters was negligible.



\subsection{Models including strong gravity}
\label{sect:fit_with}

The spectral fits detailed above demonstrated that the
Fe emission line was significantly resolved, implying emission from
near the SMBH. These did not directly address the issue of a
highly redshifted component to the line. Furthermore, \mcg\ shows a 
strong reflection continuum (section~\ref{sect:ref_intro}) which needs
to be accounted for in the spectral modelling. Therefore, in this
section the $3-10\keV$ spectrum is fitted with models accounting for
emission from an X-ray illuminated accretion disc extending close to
the SMBH (see also W01, paper~I and references therein). 

The code of Ross \& Fabian (1993; see also Ballantyne \et 2001) was
used to compute the spectrum of emission from a photoionised accretion
disc. The input parameters for the model were as follows: the photon
index ($\Gamma$) and normalisation ($N$) of the incident continuum,
the relative strength of the reflection compared to directly observed
continuum  ($R$) and the ionisation parameter ($\xi = 4 \pi
F_{\mathrm{X}}/n_{\mathrm{H}}$). This model computes the emission
(including relevant Fe \ka\ lines), absorption and Compton scattering
expected from the disc in thermal and ionisation equilibrium.

To account for Doppler and gravitational effects, the disc emission
spectrum was convolved with a relativistic kernel calculated using the
Laor (1991) model. 
The calculation was carried out in the Kerr metric appropriate
for the case of a disc about a maximally spinning (`Kerr') black hole.
The model uses a broken power-law to approximate
the radial emissivity (as used in paper~I).  The parameters of the kernel
were as  follows: inner and outer radii of the disc ($r_{\rm in},
r_{\rm out}$), inclination angle ($i$), and three parameters
describing the emissivity profile of the disc, namely the indices
($q_{\rm in}, q_{\rm out}$) inside and outside the break radius
($r_{\rm br}$). In addition, narrow Gaussians were included at
$\approx 6.4\keV$ to model a weak, unresolved emission line and at
$\approx 6.7\keV$ to model the unresolved absorption feature. This
emission spectrum (power-law, reflection, narrow emission  line) was
absorbed using the model discussed in section~\ref{sect:wa_intro}.

\begin{figure}
\scalebox{0.475}{\includegraphics{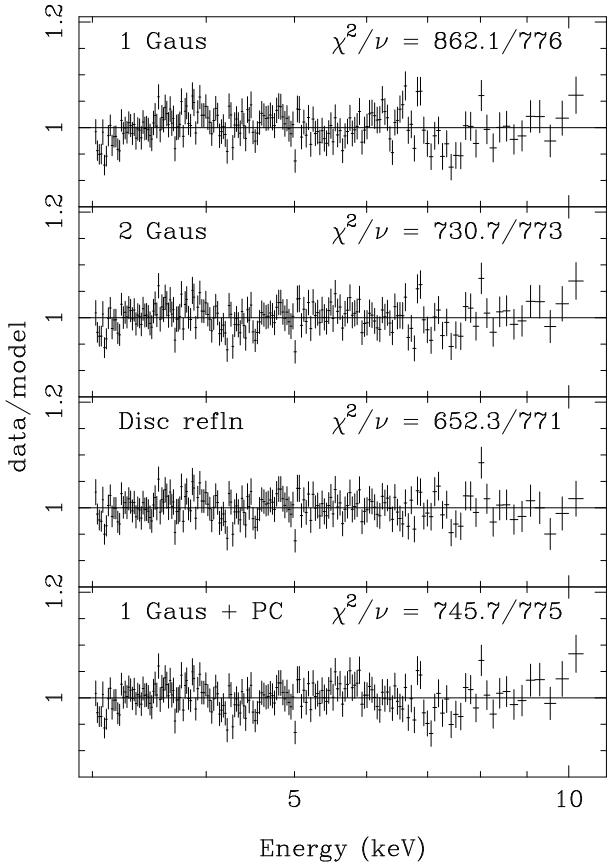}}
\caption{
Residuals from fitting the EPIC spectra with
a variety of models (for clarity only the EPIC pn:s data
are shown). The models comprise an absorbed power-law continuum plus:
($a$) broad Gaussian emission line,
($b$) two broad Gaussians,
($c$) relativistically blurred emission from an accretion disc and
($d$) broad Gaussian and partially covered continuum.
Also included is all these models are an unresolved absorption feature at
$\approx 6.7\keV$ and an unresolved, neutral iron emission line at $6.4\keV$.
}
\label{fig:comp_fits}
\end{figure}

This model gave an excellent fit to the data ($\chi^{2} = 652.3 / 771
~ dof$). The residuals are shown in Fig.~\ref{fig:comp_fits}.  The
best-fitting parameters corresponded to strong reflection
($R=1.48_{-0.06}^{+0.31}$) from a weakly ionised disc ($\log(\xi) \ls
1.4$).  The parameters of the blurring kernel defined a relativistic
profile from a disc extending in to $r_{\rm in} = 1.8 \pm 0.1 \rg$ 
with a steep (i.e. very centrally focused) emissivity function
within $\approx 3 \rg$ and a flatter emissivity
without. Specifically, the emissivity parameters were $q_{\rm in} =
6.9\pm0.6$, $q_{\rm out} = 3.0\pm0.1$ and $r_{\rm br} = 3.4\pm0.2
\rg$ and the inclination of the disc was $i=33 \pm 1 \deg$. The
narrow emission and absorption lines at $6.4\keV$ and $\approx
6.7\keV$ had equivalent widths of $\approx 10\eV$ and $\approx -10
\eV$, respectively.  This model is very similar to the best-fitting
disc model obtained in paper~I from the MOS and \sax\ data. The only
notable difference between the two models is that the emissivity is
more centrally focused in the present model. This is primarily a consequence of
including of the EPIC pn data in the fitting, which better defined the
red wing of the line, and allowing for narrow Fe \ka\ absorption
and emission lines, which subtly altered the appearance of the line profile.

The above model includes emission from well within $6\rg$, the
innermost stable, circular orbit (ISCO) for a non-rotating
(`Schwarzschild') black hole. In order to test whether a Schwarzschild
black hole (or specifically $r_{\rm in} \ge 6 \rg$) is incompatible
with the data the ionised disc model was refitted after convolution with
a {\tt diskline} kernel appropriate for emission around such a black hole (Fabian \et
1989). The disc was assumed to be weakly ionised ($\log(\xi)=1$), the
inner radius was fixed at $r_{\rm in}=6\rg$ in the fitting, and the
emissivity was described by an unbroken power-law. This model provided
an acceptable fit to the data ($\chi^{2} = 698.9 / 775~dof$) albeit
substantially worse than the model including emission from within $6
\rg$ ($\Delta \chi^{2} = 46.6$ between the two models).  Thus a model
in which the broad iron line and reflection spectrum originate from an
accretion disc about a Schwarzschild black hole cannot be ruled out
based on the $3-10\keV$ EPIC spectrum.  However, the model including
emission down to $r_{\rm in} \approx 1.8 \rg$ was preferred for two
reasons. Firstly it provided a much better fit to the data. Secondly
the derived strength of the reflection spectrum for the $r_{\rm in}
\approx 1.8 \rg$ model ($R \sim 1.4-1.8$) was more compatible with
that derived  from the independent analysis of the \sax\ PDS data
($R\gs 2$; section~\ref{sect:ref_intro}). The best-fitting value for
the Schwarzschild model was only $R=0.77\pm0.04$.

\subsection{Models excluding strong gravity}
\label{sect:fit_without}

The previous section discussed models in which the spectrum is shaped
by strong gravity effects. In this section some trial spectral models
are discussed that attempt to explain the spectrum without recourse to
strong gravity. In particular, the core of the iron line emission is
modelled in terms of a resolved Gaussian and an unresolved emission
line, both centred on $6.4\keV$ (see
section~\ref{sect:line_core}). This approximates the distortion
on the Fe emission line due to Doppler but not
gravitational effects. The asymmetry
around the line, in particular the broad excess on the red side, is
instead modelled in terms  of additional emission components rather than
allowing for highly redshifted iron line emission. The possibilities
explored are as follows:  (i) a blackbody,  (ii) a blend of broadened
emission lines from elements other than Fe, and (iii) a partially covering
absorber.

\begin{enumerate}

\item
Blackbody emission could provide a broad bump not dissimilar from
an extremely broad line. A model comprising a power-law, three
Gaussians (resolved and unresolved Fe emission lines and the
unresolved absorption feature at $\approx 6.7\keV$) and a blackbody
gave a reasonable fit to the data ($\chi^{2}=767.0/775~dof$). The
blackbody temperature was $kT\approx 1.1\keV$ and it produced a broad
excess in flux from $<2\keV$ to $\sim 6\keV$, i.e. on the red side of
the line.  This model is therefore formally acceptable but the fit is
considerably worse than with the relativistic disc models.

\item
Along with the \ka\ emission from Fe there may also be \ka\ emission
from e.g. S, Ar, Ca or Cr that would also be Doppler broadened. An
additional broad Gaussian line was included in the model with a
centroid energy in the range $3-5.5\keV$, and a width fixed to be the
same as the resolved Fe line.  This provided an unacceptable fit
($\chi^{2}=920.6/780~dof$).  Adding a third broad Gaussian yielded
an acceptable fit ($\chi^{2}=809.7/778~dof$). The additional lines
were centred on $\approx 3.7\keV$ and $\approx 5.0\keV$.  The first of
these could plausibly be due to Ca~\textsc{xix-xx}, but may be caused by 
a calibration artifact (see section~\ref{sect:nh_test}).
No strong X-ray line is
expected at $\approx 5.0\keV$.  The line equivalent widths were
$\approx 105\keV$ and $\approx 60\keV$, at least an order of magnitude
higher than expected for fluorescence from low-$Z$ elements (e.g. Matt
\et 1997; Ross \& Fabian 1993).

\item
The presence of partially covered emission (i.e. a patchy absorber;
Holt \et 1980) can cause `humps' in the $2-10\keV$ spectral range.  In
such model the observed spectrum is the sum of absorbed and unabsorbed
emission. A neutral partial covering absorber (with column density,
$\nh$, and covering fraction, $f_{\rm c}$, left as free parameters)
was applied to the power-law continuum in the spectral model. Such a
model can provide a reasonable fit to the data
($\chi^{2}=745.7/775~dof$) with $\nh \approx 1.4\times 10^{23}
\pscm$ and $f_{\rm c}=0.27\pm0.03$.

\end{enumerate}

This analysis demonstrated that alternative emission components can
reproduce the broad spectral feature observed in the $3-10\keV$ EPIC
spectrum.  However, these models all give best-fitting $\chi^{2}$
values considerably worse than the relativistic disc model discussed
in section~\ref{sect:fit_with}. Possibilities (i) and (ii) are rather
ad hoc and physically implausible. A partially covering absorber seems
somewhat more reasonable since partial covering has been claimed for
other Seyfert galaxies. However, such a model cannot explain the high
energy data, in particular, the need for a strong reflection component
(section~\ref{sect:ref_intro} and see also Reynolds \et
2003). Furthermore, allowing for partial covering in the model did not
eliminate the need for a strongly Doppler broadened iron line.

As a final test of the possible impact of partial covering on fitting
the relativistic disc models, the model from
section~\ref{sect:fit_with} was re-fitted including  partial covering.
The addition of a partially covering absorber to the best-fitting
reflection model did not provide a significant improvement to the fit
($\Delta \chi^{2} = 2.6$ for two additional free parameters). The
derived covering fraction of the absorber was poorly constrained
($0.21 \le f_{\rm c} \le 1.0$) with a column density  $\nh =
5.1_{-0.4}^{+4.9} \times 10^{21} \pscm$. The inclusion of the partial
coverer did not substantially alter the derived values of the
relativistic profile parameters (specifically $r_{\rm in} = 1.8 \pm
0.1\rg$).  This analysis suggests that the possible presence of
partially covering absorption does not  alter the derived parameters
for the relativistic emission line.  Furthermore, the inclusion of the
partial coverer resulted in a lower value  for the reflection strength
($R \approx 1.1$), in contradiction with the PDS spectrum ($R\gs 2$
see section~\ref{sect:ref_intro}).



\section{Spectral variability}
\label{sect:spec_var}

The rapid X-ray variability of \mcg\ has long been known to 
show complex energy dependence (e.g. Nandra, Pounds \& Stewart 1990;
Matsuoka \et 1990; Lee \et 1999; Vaughan \& Edelson 2001). Vaughan \et
(2003a) used 
Fourier methods to examine differences between broad energy bands
as a function of timescale. In this section the spectral variability
is examined in a finer energy scale, at the expense of cruder
timescale resolution, in order to probe the details of changes due
to specific spectral components. The EPIC 
data were used over the $0.2-10\keV$ band as the calibration issues
discussed above (see also Appendix~\ref{app:cal}) will not affect
these analyses. 

\subsection{Flux-flux analysis}
\label{sect:ff}

In this section the spectral variability is investigated using
`flux-flux' plots.  Taylor, Uttley \& M$^{\rm c}$Hardy (2003) used the
\xte\ data of \mcg\ to investigate the correlations between the fluxes
in different energy ranges. This section parallels their analysis.

\begin{figure}
\rotatebox{270}{ \scalebox{0.35}{\includegraphics{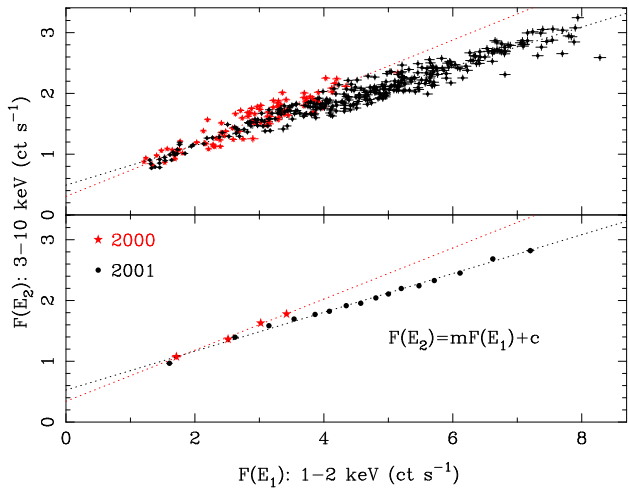}}}
\caption{
Top panel: correlation diagram comparing EPIC pn count rates
in two different energy bands. The points are from $1000\s$ binned light
curves. The two observations (2000 and 2001) are marked with different
symbols. The dashed line represents the best fitting linear function.
Bottom panel: correlation diagram after averaging the data in bins of
$20$ points. 
The points are shown with error bars on both axes, which are
of comparable size to the points.
The dashed line represents the best fit to the binned data.
}
\label{fig:ff}
\end{figure}

Figure~\ref{fig:ff} shows the count rates in a soft band ($1-2\keV$)
against the simultaneous count rates in a hard band ($3-10\keV$) from
the $1000\s$ binned EPIC pn light curves.  Clearly the two sets of count
rates (hereafter fluxes) show a very significant correlation. This is
true for both \xmm\ observations.  The data were fitted with a linear
model (of the form $F(E_2) = m F(E_1) + c$; where $F(E_1)$ and
$F(E_2)$ represent 
the fluxes in the two bands) accounting for errors in both axes. This
resulted in a formally unacceptable fit (rejection probability
$>99.99$ per cent).  However, the poorness of the fit is a
result of intrinsic scatter in the correlation, rather than any
non-linearity. This intrinsic scatter in the relation between
different energy ranges is a manifestation of both the difference in
shape of the power spectral density (PSD) function and low coherence
between variations in different 
energy bands (Vaughan \et 2003a).  Therefore, to overcome this the data
were then binned such that each bin contains $20$ points and errors were
calculated in the usual fashion (equation 4.14 of Bevington \&
Robinson 1992). The binned data, which reveal the average flux-flux
relation, provided a reasonable fit to a linear
function (rejection probabilities of  $33$ and $89$ per cent for the
2000 and 2001 datasets, respectively).

As concluded by Taylor \et (2003) the tight linearity of the flux-flux
correlation (at least over the relatively short time scales probed by
each \xmm\ observation) strongly indicates that the flux variations
are dominated by changes in the normalisation of a spectral component
with constant softness ratio $m$ (i.e. the gradient of the
flux-flux correlation). In addition there must be an additional
spectral component that varies little and contributes more in the
$3-10\keV$ band than the $1-2\keV$ band and therefore produces the
positive constant offset $c$. In other words the flux in each band is
the sum of a variable component and a constant component: 
\begin{equation}
F(E) = N_{\rm V} V(E) + N_{\rm C} C(E)
\label{eqn:2comp}
\end{equation}
(where $N_{\rm V}$ and $N_{\rm C}$ are the normalisations of
the variable and constant spectra $V(E)$ and $C(E)$ respectively)
\footnote{The spectra of the two components $V(E)$ and $C(E)$ include
the effects of absorption.}.
The fact that the relation between $F(E_1)$ and $F(E_2)$ is linear
means that the gradient gives $m=V(E_2)/V(E_1)$ and the offset gives
$c=C(E_2) - mC(E_1)$.
As can be seen from the figure the
relation changed between the 2000 and the 2001 observations; these
changes in  slope and offset imply the shape of the two spectral
components differed slightly between the two \xmm\ observations.
The flux-correlated changes in the spectrum are caused by changes
in the relative contributions of these two spectral components. 
This `two component' interpretation of the X-ray spectral variability
of Seyfert 1s has been discussed by e.g. Matsuoka \et
(1990); Nandra (1991); M$^{\rm c}$Hardy \et (1998); Shih \et (2002);
Fabian \& Vaughan (2003).

\subsubsection{Decomposing the spectrum}

\begin{figure}
\rotatebox{270}{
\scalebox{0.35}{\includegraphics{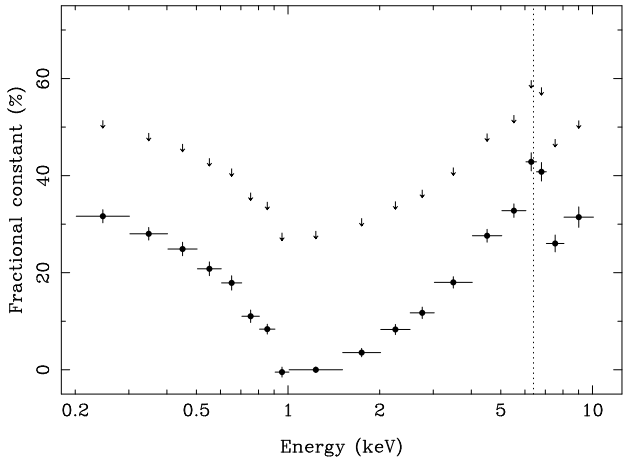}}}
\caption{
Fractional contribution to the spectrum of the constant component
($C(E)/\overline{F(E)}$ deduced from the linear flux-flux
relation). The errors are $1\sigma$  
confidence limits obtained from the linear fit to the flux-flux data.
If the constant component contributes a non-zero fraction in the
$1.0-1.5\keV$ range the contributions at all energies will increase
but the spectral shape will remain the same. The arrows indicate the
absolute upper limits on the fraction of constant emission (see text).
}
\label{fig:const}
\end{figure}

As discussed in Fabian \& Vaughan (2003), the `difference spectrum'
can be used to determine the shape of the variable component  of the
spectrum, $V(E)$ (see section~\ref{sect:fluxes}).  The offset of the
flux-flux plot, $c$, gives a measure of the strength of the constant
component in the spectrum and so by measuring the offset  as a
function of energy, $c(E)$ it is possible to estimate the spectrum of
the constant component $C(E)$ (see section 4.3 of Taylor \et 2003).
Figure~\ref{fig:const} shows the spectrum of the constant component in
\mcg\ deduced by measuring the offset $c(E)$ of the flux-flux
relations.  Binned flux-flux plots were constructed by comparing the
light curves in each energy band to the $1.0-1.5\keV$ band
($F(E_1)$). In each case the constant offset was measured from the
best fitting linear model and normalised to the average flux in the
energy band under examination.  This yielded an estimate of the
fractional contribution to the spectrum from non-varying components.

The spectrum of the constant component shown in figure~\ref{fig:const}
reveals that the constant component is soft below $\sim 1\keV$ and
hard at higher energies. The largest fractional  contribution from the
constant component occurs in the iron K band,  where it contributes
$\sim 40$ per cent of the total flux.  There are two assumptions implicit in
this analysis. The first is that the variable component has a constant
spectral shape $V(E)$ 
when averaged as a function of flux (which is implied by the
linearity of the flux-flux relations). The second assumption is that
there is a negligible 
constant component in the  $1.0-1.5\keV$ band (which was used as the
comparison), i.e. $C(E_1) = 0$. If the  $1.0-1.5\keV$ emission does
contain a non-zero constant component ($C(E_1)>0$) then the absolute
scale of the inferred spectrum would increase but its spectral form
will be approximately the same.  The maximum possible amount of
constant emission in the $1.0-1.5\keV$ band is given by the minimum flux in
that band (i.e. $0\le C(E_1) \le \min[F(E_1)]$), therefore this can be used to
place an absolute limit on the strength of the constant component:
$c(E) \le C(E) \le \{ c(E) + m(E)\min[F(E_1)] \}$. This upper
limit is also indicated in figure~\ref{fig:const}.  It should be noted
that the constant component $C(E)$ must be affected  by the soft X-ray
absorption. The lack of features associated with absorption in
figure~\ref{fig:const} (in particular the lack of a spectral jump at
$\sim 0.7\keV$) demonstrates that the constant component is affected
by absorption exactly as is the total spectrum (hence the absorption
was factored out by normalising the spectrum $C(E)$ by the total
spectrum). This spectrum can be identified with the
Reflection-Dominated Component (RDC) of Fabian \& Vaughan (2003) and is
henceforth referred to as the RDC.

\subsection{Flux-resolved spectra}
\label{sect:fluxes}

\begin{figure}
\rotatebox{270}{
\scalebox{0.35}{\includegraphics{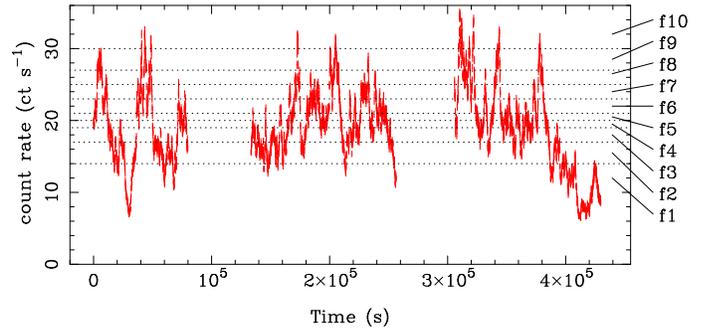}}}
\caption{
$0.2-10\keV$ EPIC pn light curve of the 2001 observation.
The light curve has been divided into ten count rate intervals marked
$f1-f10$. 
}
\label{fig:flux_lc}
\end{figure}

\begin{figure}
\rotatebox{0}{
\scalebox{0.7}{\includegraphics{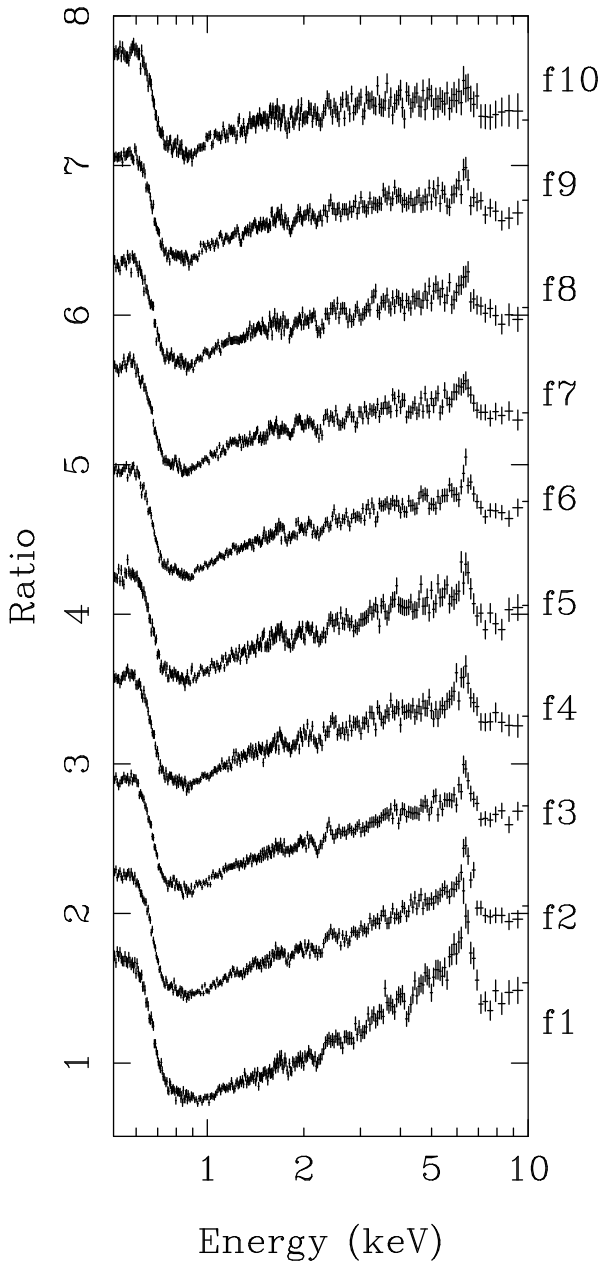}}}
\caption{
EPIC pn spectra from each of the count rate slices (shown in
figure~\ref{fig:flux_lc}) compared to a $\Gamma=2$ power-law model
(modified by Galactic absorption). The data have been shifted upwards
to aid clarity. It can be seen that as the flux decreases from $f10$
to $f1$ the overall spectrum becomes harder (above $2\keV$) and the strength 
of the iron feature decreases (relative to the continuum model).
}
\label{fig:flux_fit}
\end{figure}

\begin{figure}
\rotatebox{0}{
\scalebox{0.7}{\includegraphics{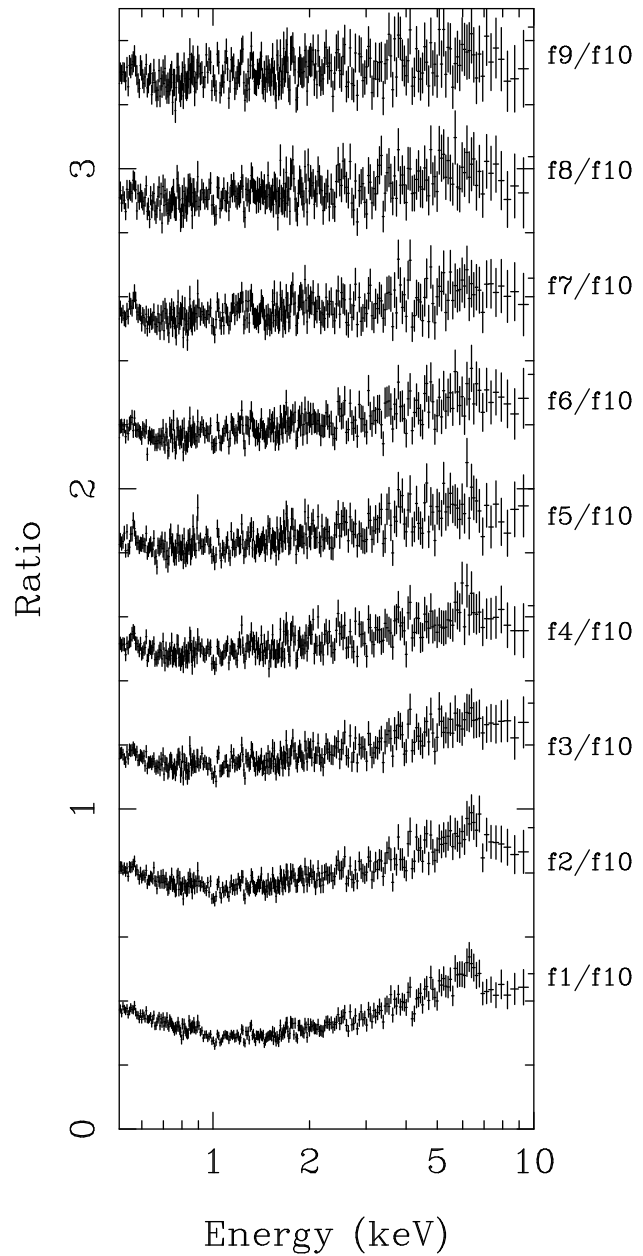}}}
\caption{
EPIC pn spectra from the count rate slices shown as a ratio
to the highest count rate spectrum ($f10$).
Similar to figure~\ref{fig:flux_fit} the spectra can be seen
to become harder and show relatively stronger iron features as
the count rate decreases.
}
\label{fig:flux_rat}
\end{figure}

A more conventional way to examine how the spectrum of the source
evolves with flux is to extract energy spectra from different
flux `slices.' This was done for the 2001 \xmm\ data by dividing
the pn data into ten flux intervals such that the total number of photons
collected from the source in each flux interval was comparable ($\sim
5\times 10^5$). 
The highest and lowest flux slices differ by a factor of $3$ in count
rate and all ten are marked on the light curve in 
Figure~\ref{fig:flux_lc}. The ten spectra extracted from each flux
slice were then binned in an identical fashion. There were clear,
systematic spectral changes between the ten flux intervals, these are
illustrated in two different ways in figures~\ref{fig:flux_fit} and
\ref{fig:flux_rat}. Figure~\ref{fig:flux_fit} shows the ten spectra
as ratios to a ($\Gamma=2$) power-law model (modified only by
Galactic absorption). This illustrates the softening of the $2-10\keV$
spectrum as the source gets brighter. In addition the strength of the
iron line relative to the continuum gets weaker as the source
brightens (i.e. the equivalent width decreases with increasing continuum
luminosity) meaning that the iron line becomes a lower contrast
feature at high fluxes.

Figure~\ref{fig:flux_rat} shows nine of the ten spectra as
a ratio to the highest flux spectrum ($f10$). This again shows the
spectrum became harder (in the $2-10\keV$ range) and the iron line
became relatively stronger as the source became fainter.
It is also interesting to note that at energies below $\sim 2\keV$ the
spectrum becomes softer as the source becomes fainter, contrary to
the trend at higher energies. In addition, there is no strong feature in the
spectral ratios at $\sim 0.7\keV$, indicating that the fractional 
strength of the $\sim 0.7\keV$ spectral jump remains constant with
flux. This is consistent with the jump being due to absorption with
a constant optical depth.

These average changes in the detailed spectral shape as a function of
flux can be explained using the simple two-component model discussed
above. The inferred constant emission comprises a soft excess below
$\sim  1\keV$, a harder tail above $\sim 2\keV$, and a strong iron
line. One would expect the total (variable$+$constant) spectrum to
change  exactly as observed: as the total flux decreases the
contribution from the constant  component, relative to the variable
component, increases. Therefore, at low fluxes, the spectrum will
become softer below 
$\sim 1\keV$,  harder above $\sim 2\keV$, and display a more
prominent iron line.

\subsubsection{Average EPIC difference spectrum}
\label{sect:diff}

The five highest flux spectra were combined, as were the five lowest
flux spectra, to produce two spectra representing the source during
flux intervals above and below the mean (hereafter `high' and `low'
fluxes). Subtracting the low flux spectrum from the high flux
spectrum will remove the contribution from the constant component.
As discussed in Fabian \& Vaughan (2003), under the 
assumption that the total spectrum can be accurately described (on
average) by the superposition of two emission components, one variable
and one not (both modified by absorption), the difference spectrum will 
give the spectrum of only the variable component (modified by
absorption): $F_{hi}(E)-F_{lo}(E) = (N_{\rm V, hi} - N_{\rm V, lo})
V(E)$ (where $V(E)$ includes absorption).

\begin{figure}
\rotatebox{270}{
\scalebox{0.35}{\includegraphics{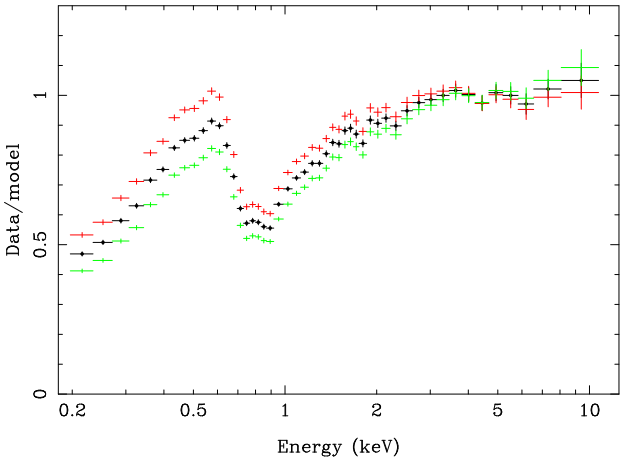}}}
\caption{
EPIC pn difference spectra produced by subtracting the
low flux spectrum from the high flux spectrum. The difference
spectrum is shown as a ratio to a power-law (modified 
by Galactic absorption) fitted across the $3-10\keV$ range.
The data/model ratio calculated using the best fitting power-law
slope is shown in black (circles). Above and below this lie the
ratios calculated assuming the $90$ per cent lower and upper limits
on the slope of the $3-10\keV$ power-law.
The data were rebinned for clarity.
}
\label{fig:diff_spec}
\end{figure}

Figure~\ref{fig:diff_spec} shows the EPIC pn difference spectrum
(compared to a power-law model fitted in the $3-10\keV$ range) from
the 2001 observation. A power-law (modified only by Galactic
absorption) gave an excellent fit to the $3-10\keV$ data
($\chi^{2}=28.1/60~dof$) and yielded a photon index of $\Gamma = 2.20
\pm 0.05$. Extrapolating this power-law down to lower energies reveals
the signature of strong absorption, although the exact amount depends
on the assumed shape of the  underlying continuum (as illustrated by
the upper and lower datasets shown in the figure). It also be noted
that if the variable spectral component steepens at low energies
(i.e. contains a soft excess) the inferred absorption profile at lower
energies could be systematically underestimated even further (see
section 6 of Turner \et 2003a).  Leaving aside these uncertainties on
the inferred spectrum at lower energies, it remains true that above
$3\keV$ the average spectrum of the varying component $V(E)$ can be
described accurately as a single power-law ($\Gamma \approx 2.2$).
In particular it should be noted that the difference spectrum (unlike
the spectrum of the constant component described above) does not
possess any obvious features in the iron K region. The upper limit
on the flux of the line core (section~\ref{sect:line_core}) present in the
difference spectrum is a factor $\sim 20$ lower than that present
in the time averaged spectrum. 

A further investigation of the flux-resolved spectra revealed
that the difference spectrum is consistent with a power-law
at all flux levels. Nine difference spectra were calculated
by subtracting the lowest flux spectrum ($f1$) from each of the 
nine higher flux spectra ($f2-f10$). These nine difference
spectra were fitted over the $3 - 10\keV$ range with an absorbed
power-law model. In all nine cases the simple power-law model
was found to provide a good fit to the data ($\chi_{\nu}^{2}\approx
1.0$), with no evidence for strong systematic residuals around the
Fe K band. The limit on the depth of possible Fe edges in the $7.1-8
\keV$ region was $\tau \ls 0.07$, consistent with the limits
obtained in section~\ref{sect:fe_test}.
In the discussion below this variable component to the
spectrum will be referred to as the variable Power-law Component
(PLC; Fabian \& Vaughan 2003).

\subsection{rms spectra}
\label{sect:rms}

\begin{figure}
\scalebox{0.475}{\includegraphics{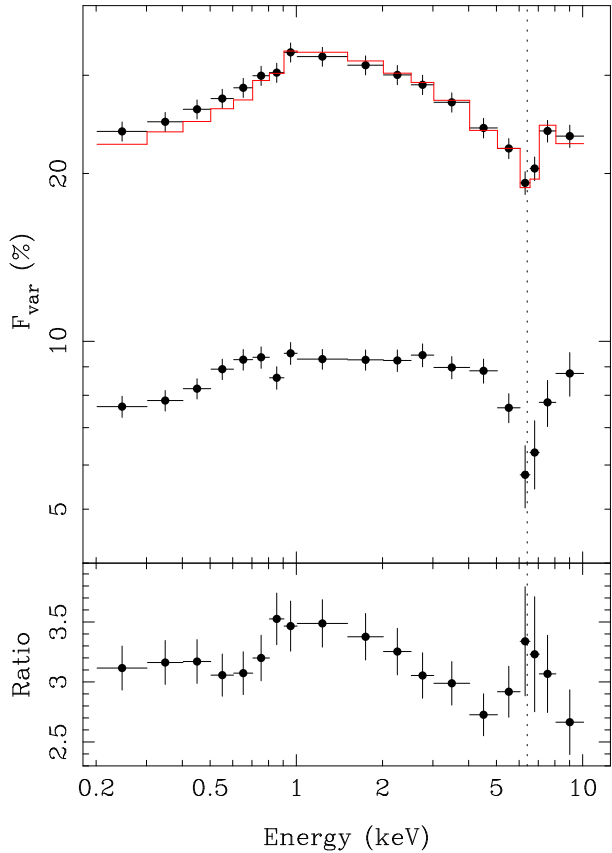}}
\caption{
Top panel: fractional rms variability amplitudes as a function
of energy on two different timescales. The upper data represent
the amplitude integrated over the entire (2001) observation while the
lower data represent the rms deviation between neighbouring points 
(point-to-point rms) and therefore samples only short timescale
variability ($\sim 1000\s$).
The errors were calculated as described in the text.
The dashed line indicates the energy of the (neutral) Fe \ka\
line ($6.40\keV$).
The histogram shows the `two component' model rms spectrum. 
Bottom panel: The ratio of the two rms spectra, showing how the energy
dependence of the variability amplitude changes with variability timescale.
}
\label{fig:rms}
\end{figure}

The root mean squared (rms) spectrum measures the variability
amplitude as a function of energy and can in  principle reveal which
spectral components are associated with the strongest
variability (see Edelson \et 2002 and Vaughan \et 2003b). 
Figure~\ref{fig:rms} shows two rms spectra calculated
from the 2001 \xmm\ observation of \mcg, clearly showing in both cases
a strong dependence of the variability amplitude on energy.

The two rms spectra show the variability amplitude on different
timescales. The upper spectrum shows the fractional rms amplitude
integrated over the entire observation in the standard fashion (eqn.~1
of Edelson \et 2002) using $1000\s$ binned light curves. As the
variability is dominated by long timescale changes (Uttley \et 2002;
Vaughan \et 2003a) this therefore reveals the energy dependence of
variations occurring on timescales comparable to the length of the
observation ($\sim 100\ks$). The lower spectrum shows the fractional
rms of the point-to-point deviation (i.e. the rms difference
between adjacent time bins, as defined by eqn.~3 of Edelson \et
2002).  This therefore only measures fluctuations between
neighbouring time bins and so probes the energy dependence of the
variability on short timescales comparable to the bin size ($\sim
1\ks$).  The errors were calculated as in eqn.~2 of Edelson \et
(2002), but as discussed in their appendix should be considered only
as approximations since they strictly assume the light curves are
drawn  from independent Gaussian processes. Matsumoto \et (2003)
previously used the longest \asca\ observation of \mcg\ to 
obtain rms spectra on different timescales.

As is clear from the figures the rms spectra show broad peak at $\sim
1\keV$ and a localised
depression around the energy of the iron K emission line, strongly
suggesting a  suppression of the variability due to the presence of
the emission line (see also Inoue \& Matsumoto 2001).  
The exact energy dependence of the variability amplitude differs
between the two timescales (evident from  the ratio of the two rms
spectra), as should be expected if the shape of the PSD
is a function of energy (Vaughan \et 2003a).

\subsubsection{Modelling the rms spectrum}

In the previous sections is was shown how the spectral variability  is
consistent with a model comprising a variable component $V(E)$ (the PLC; see
Figure~\ref{fig:diff_spec}) and a constant component $C(E)$ (the RDC; see
Figure~\ref{fig:const}). Given these empirically derived spectral
components it is possible to construct a model for the rms spectrum
shown in Figure~\ref{fig:rms} (see also Inoue \& Matsumoto 2001).

It is assumed that the spectrum is given by equation~\ref{eqn:2comp}
where both $V(E)$ and $C(E)$ include absorption effects. The
(normalised) rms spectrum can be expressed as:
\begin{equation}
F_{\rm var} (E) = \frac{ \sigma [F(E)] }{ \overline{F(E)} } =
\frac{ \sqrt{ \sigma[N_{\rm V} V(E)]^2 + \sigma[N_{\rm C} C(E)]^2 } }{
  \overline{N_{\rm V}} V(E) + \overline{N_{\rm C}} C(E) }
\label{eqn:fvar_model}
\end{equation}
where $\sigma[F(E)]$ and $\overline{F(E)}$ represent
the absolute rms amplitude and time-average of the spectrum 
over the observation, respectively. (Likewise for the two spectral
components $V(E)$ and $C(E)$.) In terms of the model
being discussed it is assumed that $C(E)$ is not variable and so
$\sigma[ N_{\rm C} C(E)] =0$. Therefore:
\begin{eqnarray}
F_{\rm var} (E) = 
\frac{ \sigma[N_{\rm V} V(E)] }{ \overline{N_{\rm V}} V(E) +
  \overline{N_{\rm C}} C(E)} 
\nonumber \\
=\frac{ \sigma[ N_{\rm V} ]}{\overline{N_{\rm V}}}  ( 1 - C(E) /
\overline{F(E)} )   
\label{eqn:fvar_model2}
\end{eqnarray}
and the term $C(E)/\overline{F(E)}$ is the fractional contribution
of the constant component shown in Figure~\ref{fig:const}. 

The above analysis assumes that the variable component changes only in
normalisation  and not in shape (i.e. $\sigma[ N_{\rm V} V(E)] /
\overline{N_{\rm V} V(E)} = \sigma[ N_{\rm V} ]/  \overline{N_{\rm
V}}$), but is independent of the actual form of the spectrum $V(E)$.
The linearity of the flux-flux relations implies the variable
component  has a constant softness ratio with flux, and therefore its
spectrum $V(E)$ is independent of flux, validating this assumption.
The first term on the right-hand side of
equation~\ref{eqn:fvar_model2} is thus simply a normalisation.
Implicit in this analysis is the assumption that the absorption
function is not time-variable. If the absorption function did vary
this would introduce another term to the rms spectral model (Inoue \&
Matsumoto 2001), but such a term is not required by the data and thus
it was assumed that the absorption did not vary.

The histogram in Figure~\ref{fig:rms} shows the resulting rms spectral
model when the constant component derived in section~\ref{sect:ff}
(Figure~\ref{fig:const}) is used. This very simple model clearly
reproduces (to within $\sim 1$ per cent in $F_{\rm var}$) the energy
dependence of the variability amplitude on these timescales. The peak
at $\sim 1\keV$ is a result of the constant component $C(E)$ being
weakest at $\sim 1\keV$ and the suppression at $\sim 6.4\keV$ results
from the strong iron line present in the constant component. The short
timescale rms spectrum has a subtly different shape and thus it not so
well reproduced using this simple model. However, it is known that at
high frequencies the variability  in different energy bands shows
different PSD shapes and incoherent variability. Thus the short
timescale rms spectrum will be affected by these additional effects,
although their physical origin is not clear.
This could perhaps be explained if the RDC (or perhaps the warm
absorber) varies on short timescales but these changes are
`washed out' over longer timescales. 

\subsubsection{A low-flux rms spectrum}

The shape of the rms spectrum is clearly a function of timescale, as
expected given the energy-dependent PSD (Vaughan \et 2003a). 
It also shows subtle changes with time when the rms spectrum
of each revolution of \xmm\ data are compared (an rms spectrum for the
first revolution was shown in paper~I). These may be caused
by subtle changes in the RDC spectrum. As the relative strength of the RDC
should be highest when the overall source flux is low, 
the low flux intervals of the light curve are the most promising to
search for changes in the RDC.

\begin{figure}
\scalebox{0.475}{\includegraphics{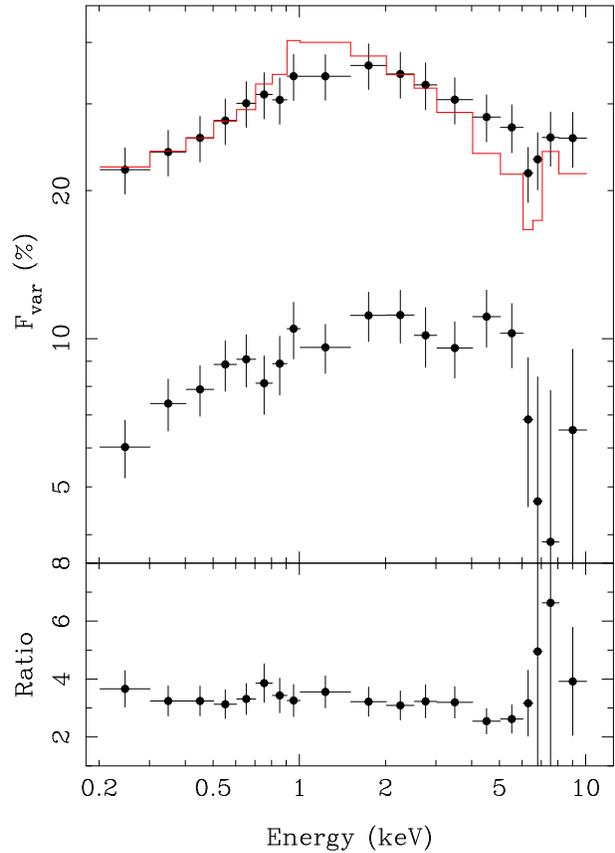}}
\caption{
The rms spectra, as shown in Fig.~\ref{fig:rms}, including
only the low flux interval towards the end of the observation.
}
\label{fig:low_rms}
\end{figure}

The rms analysis described above was repeated using only the last $45\ks$ of 
data from rev 0303. During this time interval the source flux was
rather low and in fact reached its minimum for the observation
(Fig.~\ref{fig:flux_lc}).  
The rms spectra from this interval are shown in
Fig.~\ref{fig:low_rms}. The low-flux rms spectrum, particularly on the
longer timescale, shows a similar overall shape to the rms spectrum
from the entire observation (upper data from Figs.~~\ref{fig:rms}
and ~\ref{fig:low_rms}). Interestingly the prediction of the 
two component model discussed above does not match these data so
well. In particular the model predicts either too little rms around
the iron line or too much in the soft X-ray band. 
This could be indicating that at low fluxes some variation in the
shape/strength  
of the RDC were detected (see also Reynolds \et 2003). 
However, this could also result from 
a random effect caused by the energy-dependent PSD (Vaughan
\et 2003a,b). More observations at very low flux levels could
confirm whether the RDC shows rapid variability.

\subsection{Time-resolved spectra}
\label{sect:time-res}

\begin{figure*}
  \scalebox{1.03}{\includegraphics{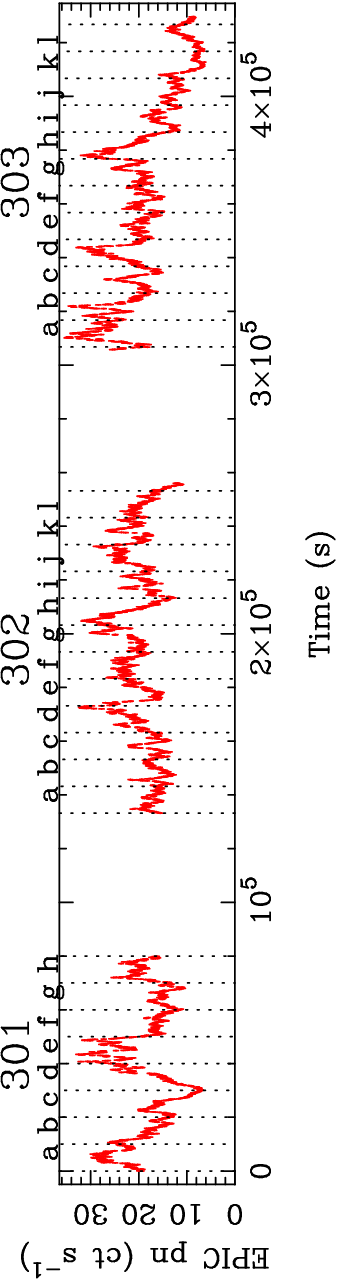}}
 \hspace{0.7 cm}
\rotatebox{90}{
\hbox{
 \scalebox{0.45}{\includegraphics{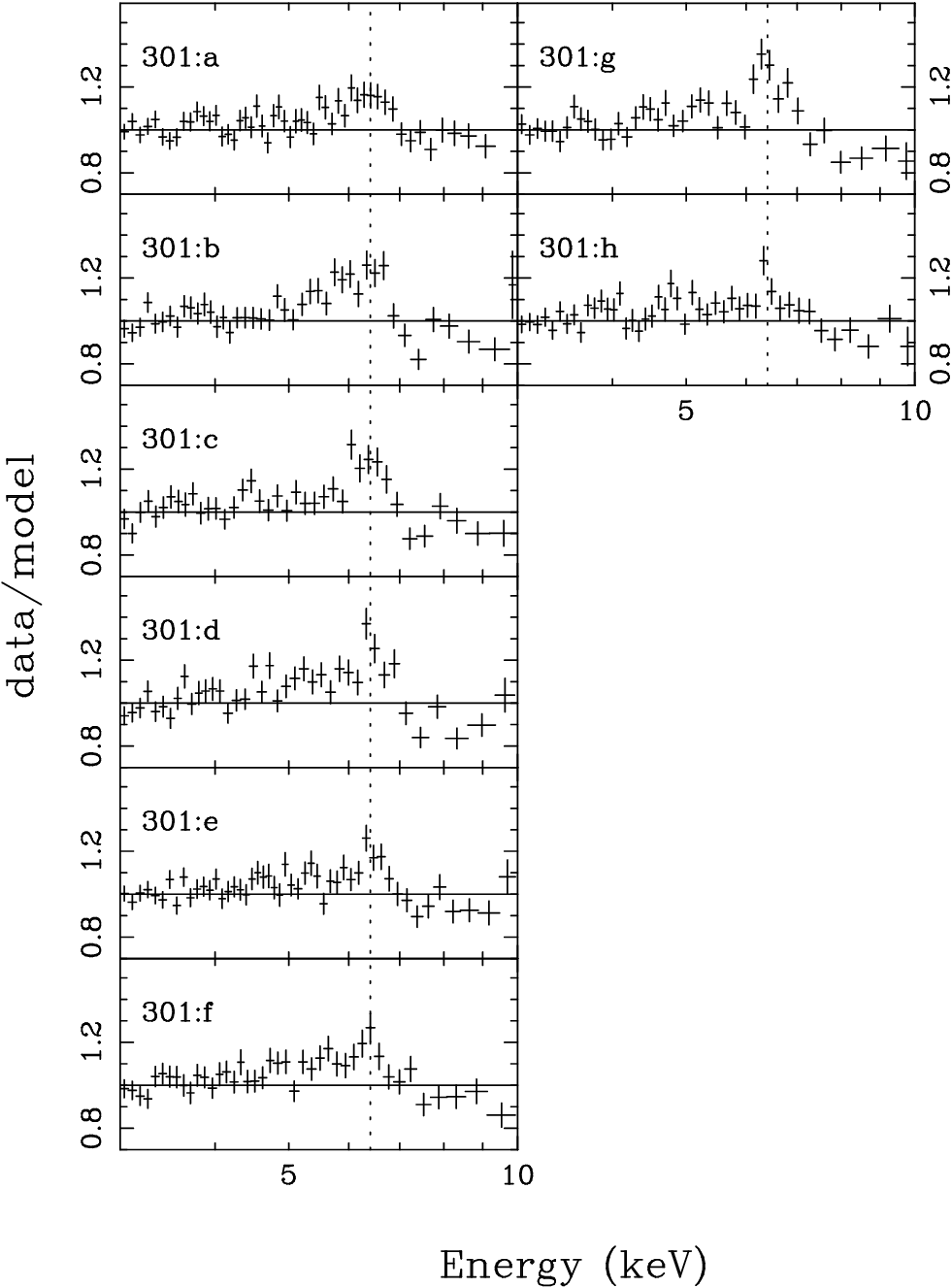}}
 \scalebox{0.45}{\includegraphics{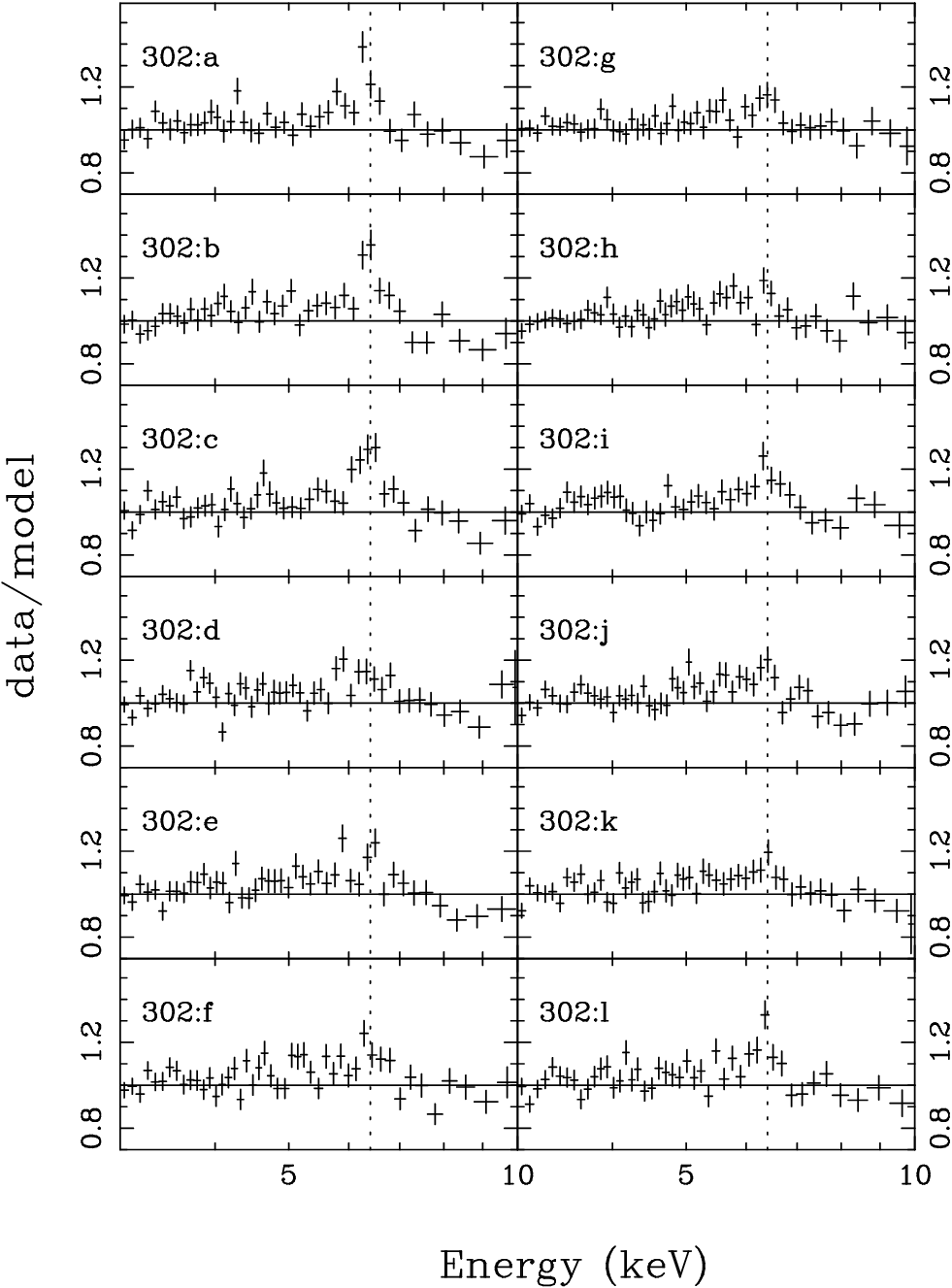}}
 \scalebox{0.45}{\includegraphics{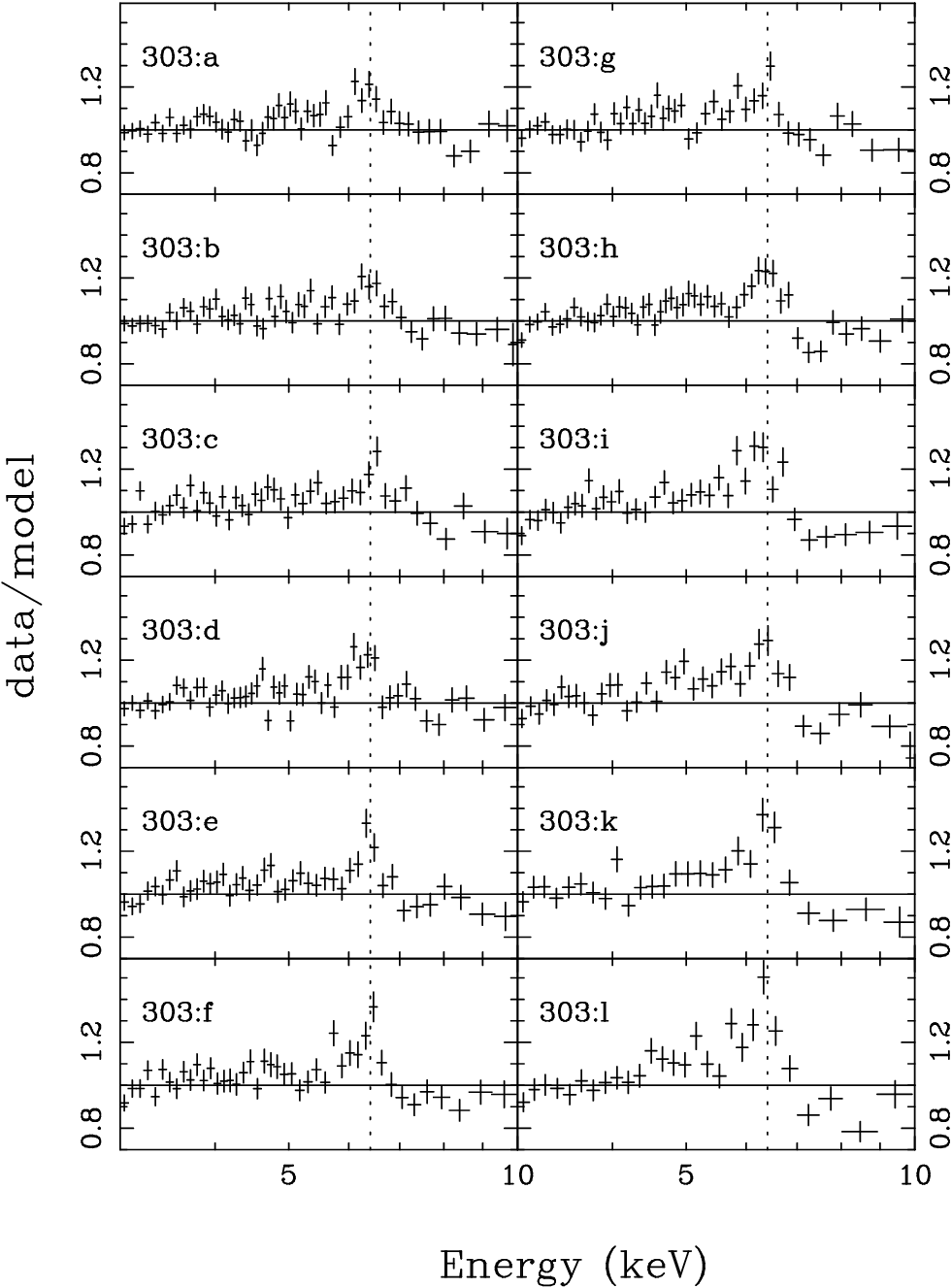}}
}
}
\caption{
Top panel: EPIC pn $0.2-10\keV$ light curve spanning all three
\xmm\ revolutions with $10\ks$ time intervals labelled.
Bottom panels: EPIC pn spectra extracted from the $10\ks$
time intervals, shown as a ratio to a simple power-law model. 
The expected energy of the iron line ($6.4\keV$) is marked
with a dotted line.
}
\label{fig:time_slices}
\end{figure*}

For completeness the spectrum was examined in consecutive
$10\ks$ time slices. Fig.~\ref{fig:time_slices} shows
the spectra extracted from each of these time slices.
Clearly the strength of the iron line (relative to the
continuum model) changes throughout the light curve.
However, the only obvious, visible change is 
that described above; as the overall X-ray 
flux decreases the iron line becomes more prominent
because it varies less than the continuum. 
For example, the spectrum for segment 303:l, which has one of the 
lowest fluxes, also has the most prominent line.

It is possible that there are short-lived spectral
features that would be `washed out' in the time-averaged
spectrum. Turner, Kraemer \& Reeves (2003b) discuss
the possible existence of transient, redshifted iron lines.
However, given that $32$ independent spectra were examined here, 
the likelihood of an apparently significant feature appearing purely
by chance is quite high. For example, the probability of detecting
a feature, considered detected at $99.7$ per cent
confidence in one individual spectrum, 
in one of the $32$ spectra is $\sim 0.1$ (using $P_N \approx 1
- (1 - P_1)^N$). By the same argument only features found at
$>3.5\sigma$ significance in any individual spectrum should be
considered as significant detections from the entire dataset
(i.e. probability of false detection $P_N \ls 0.01$).
There are no clear examples of strong, sharp
features in these data except for the obvious iron line.
The analyses of Fabian \& Vaughan (2003) and Ballantyne \et (2003)
demonstrated that these time-resolved spectra can be 
fitted using relatively simple reflection models.
In addition, there were no sharp features seen in the 
rms spectra (section~\ref{sect:rms}) that would indicate
a preferred energy for transient features. 

\subsection{Principal Component Analysis}
\label{sect:pca}

The method of Principal Component Analysis (PCA) was
used to try and isolate independently varying components present in the
time-resolved spectra discussed above.  PCA is a powerful statistical
tool used widely in the social sciences where it is often known as
factor analysis. In practice PCA gives the eigenvalues and eigenvectors
of the correlation matrix of the data. The data can be thought of as an
$m \times n$ matrix comprising $m$ observations of the $n$ variables, in
which case the correlation matrix would be a symmetric $n \times n$
matrix.

The eigenvectors can be thought of as defining a new coordinate system, in the
$n$-dimensional parameter space, which best describes the variance in
the data.  The first principal component, or PC1 (the eigenvector with
the highest eigenvalue), marks the direction through the parameter
space with the largest variance.  The next Principal Component (PC2)
marks the direction with the second largest amount of variance.  The
motivation behind PCA is to extract the (hopefully few) dominant 
correlations from a complex dataset. Francis \& Wills (1999) provides a
brief introduction to PCA as applied to quasar spectra, while Whitney
(1983) and Deeming (1964) illustrate more general applications of PCA
in astronomy.

If the variance in the spectra is caused by only
a few independently varying spectral components then PCA should
reveal them in its eigenvectors. The application of PCA
to spectral variability of AGN was first discussed by
Mittaz, Penston \& Snijders (1990). The first few Principal
Components (those representing most of the variance in the data)
should reveal the shape of the relevant spectral components.
The weaker Principal Components might be expected to be dominated
by the photon noise in the spectra (which should be uncorrelated
between each of the $n$ energy bins in the $m$ spectra and so should
not distort the shape of the first Principal Components).

\begin{figure}
{\scalebox{0.47}{\includegraphics{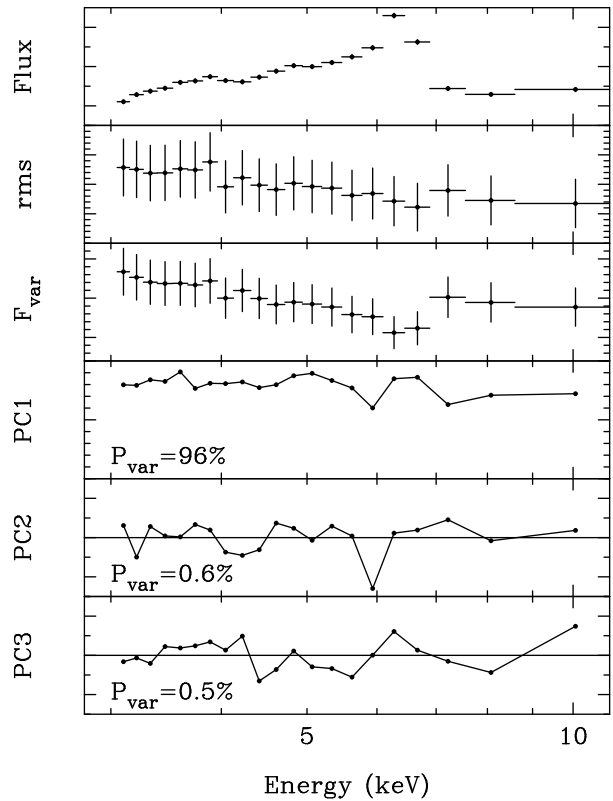}}}
\caption{
Results of Principal Component Analysis (PCA).  The top panel shows
the mean spectrum (converted into $EF(E)$ flux units).  The next two
panels show the rms and fractional rms of the data (after subtracting
the contribution from photon noise). The bottom three panels show the
spectra of the first three Principal Components and their
contribution to the total variance of the data ($P_{\rm var}$).  
The solid line marks
the zero level;  points that lie on the same side of the zero line
(all positive or all negative) are correlated with one another while
points that lie on different sides of the zero line  (differing signs) 
are anti-correlated.
}
\label{fig:pca}
\end{figure}

This method was used on the $m=32$ time-resolved spectra over the
$3-10 \keV$ range (binned such that they each spectrum contained the
same $n=21$ energy bins). The spectra were unfolded using a $\Gamma=2$
power-law model to convert them from counts to flux units (in $EF(E)$
form). In the absence of sharp features in either the source spectrum
or the detector response this should provide a reasonable estimate of
the `fluxed' data. The first three Principal Components are shown in
Figure~\ref{fig:pca} along with the mean and rms spectra derived from
the same data.

As can be clearly seen, PC1 has a relatively flat spectrum  and
describes the vast majority of the variance in the data ($96$ per
cent).  The first Principal Component can therefore be identified with
the variable PLC discussed above. All the remaining  Principal
Components each describe less than $1$ per cent  of the variance and
most likely represent only the photon noise in the data. Thus the PCA
confirms the above analyses and suggests the spectral variations in
\mcg\ are a result of a single, varying continuum component (the
PLC). Nearly identical results were found when the ranked data was
used to produce the correlation matrix (which then represented the
matrix of rank-order correlation coefficients).

\section{The 2001-2000 difference spectrum}
\label{sect:big_diff}

\begin{figure}
\rotatebox{270}
{\scalebox{0.32}{\includegraphics{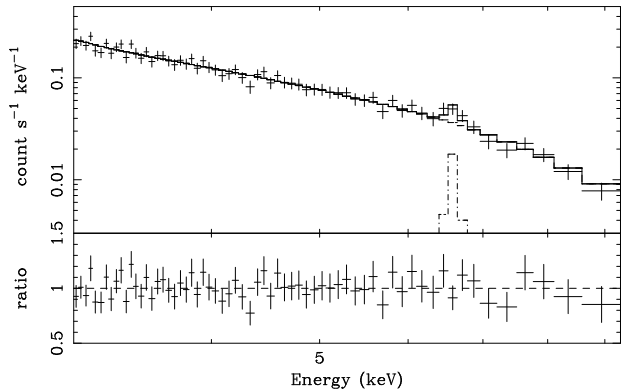}}}
\caption{
Difference spectrum produced by subtracting the 2000
data from the 2001 data. Only the EPIC pn:s spectrum 
is shown.
The only significant residual (compared to the simple
power-law continuum model) is 
an unresolved `emission' feature at $E=6.65\pm0.05\keV$.
}
\label{fig:big_diff}
\end{figure}

\begin{figure}
\rotatebox{270}
{\scalebox{0.35}{\includegraphics{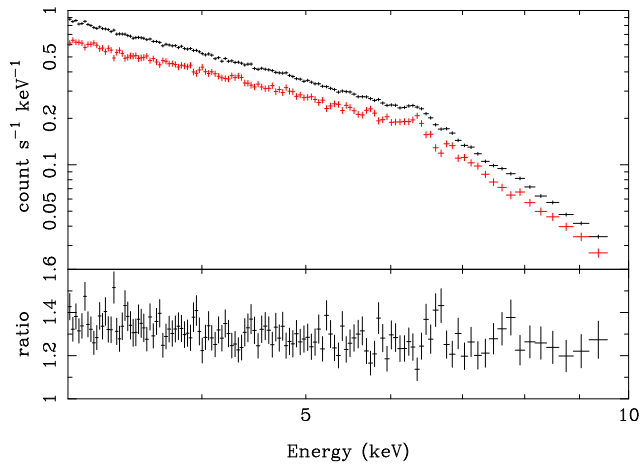}}}
\caption{
Ratio of the EPIC pn:s spectral from the 2001 (upper data)
and 2000 (lower) observations. 
The top panel shows the raw spectra, the bottom panel shows
their ratio.
This again shows the unresolved `emission' feature at $E\approx 6.65\keV$.
}
\label{fig:big_rat}
\end{figure}

Figure~\ref{fig:big_diff} shows the difference spectrum  produced by
subtracting the $3-10\keV$ EPIC pn:s spectrum taken in 2000 from the
spectrum taken in 2001. A power-law provided a good fit ($\chi^2 =
51.7/62~dof$) with $\Gamma = 2.18\pm0.09$ (compare with the results
of section~\ref{sect:diff}). The only significant residuals appeared as
an excess at $\approx 6.6\keV$. Including a narrow Gaussian improved
the fit ($\chi^2 = 44.1/60~dof$), with a best-fitting energy
$E=6.63\pm0.12\keV$. Including an absorption edge in the model did
not significantly improve the upon this fit. 
Figure~\ref{fig:big_rat} shows the ratio of the two spectra,
confirming that the two spectra differ around $E\approx 6.65\keV$.

An unresolved emission feature in the difference and ratio spectra
could be due to a change in either the flux of the broad Fe emission
line or the depth of the Fe \ka\ absorption.  In order to produce an
apparent emission feature in the first case only the blue wing of the
emission line must be stronger in the 2001 observation. In the second
case the optical depth of the absorption must be deeper in the 2000
observation. An examination of the two spectra individually suggested
the latter is more likely.  A close examination of changes in the RGS
spectrum between the two observations would clarify this issue.

\section{Review of \xmm\ results}
\label{sect:review}

In this section the main results obtained
from the long \xmm\ observation of \mcg\ are summarised
prior to the discussion in the following section.

\begin{itemize}

\item
The two \xmm\ observations (taken in 2000 and 2001) sampled
fairly typical `states' of the source. The former observation 
sampled a period of lower flux than the latter. However, the
long-term \xte\ monitoring shows this can be attributed to short-term
variability and does not imply a systematic difference between the
states of the source during the two observations
(section~\ref{sect:history}). 

\item
The high energy spectrum obtained from \sax\ shows a strong
Compton-reflection signature (paper~I; Ballantyne \et 2003), as
did the earlier \xmm/\xte\ observation (W01; Reynolds \et 2003).
There was no evidence for a low energy cut-off or roll-over in the
continuum out to $\sim 100\keV$.

\item
The RGS spectrum shows complex absorption by 
\ovii, \oviii\ and \fei\ as well as a range of other ions
(Sako \et 2003; Turner \et 2003a). The opacity is concentrated
mainly below $\sim 3\keV$ but still has an important effect on the
$3-10\keV$ spectrum (section~\ref{sect:wa_intro}). See also Lee \et (2001).

\item
The fluorescent iron line is strong and broad (section~\ref{sect:line_core}). 
The bulk of the line flux is resolved with EPIC. 
The emission peak concentrated around $6.4\keV$ is resolved with a
width $FWHM\approx 4.5\times 10^4$~\kmps, strongly indicating an origin
within $\ls 100\rg$. There is also a significant, asymmetric extension
to lower energies that indicates strong gravitational redshifts.
In addition there is a weak, intrinsically narrow core to the line emission
(sections~\ref{sect:core_intro} and \ref{sect:line_core}; see also
Lee \et 2002). 

\item
The best-fitting model for the $3 - 10\keV$ EPIC spectrum 
explained the iron emission as from the surface of a relativistic accretion disc.
The strong reflection explains the strength of both the iron line
and the Compton reflection continuum. The best fitting model 
includes emission down to $\approx 1.8\rg$
(section~\ref{sect:fit_with}; W01; paper~I). 

\item
The variations in X-ray luminosity show many striking similarities
with those seen in GBHCs such as Cygnus X-1 (Vaughan \et 2003a).  
In particular, the Power Spectral Density (PSD) function 
is similar to that expected by simply re-scaling the high/soft state 
PSD of Cyg X-1. The continuum variations are energy dependent and
show the PSD is a function of energy and that the hard variations are
delayed with respect to the soft variations. Similar results
have been found in other Seyfert 1s (NGC~7469, Papadakis, Nandra \&
Kazanas 2001; NGC~4051, M$^{\rm c}$Hardy \et 2003). 

\item
Previous observations with \asca\ (Iwasawa \et 1996, 1999; Shih \et
2003) and \xte\ (Lee \et 2001; Vaughan \& Edelson 2001) showed the
photon index of the $2-10\keV$ continuum to be correlated with its flux. 
The \xmm\ observations confirm this and demonstrate the trend
is reversed below $\sim 1\keV$, where the spectrum hardens with 
increasing flux (section~\ref{sect:fluxes}). The average
variability amplitude is highest in the range $\sim 1 - 2 \keV$, and
is lowest at energies around the iron line (section~\ref{sect:rms}). 

\item
The variable spectrum can be decomposed into two components, 
a variable Power-law component (PLC; section~\ref{sect:diff}) and a
relatively constant 
Reflection Dominated Component (RDC; section~\ref{sect:ff}). The
spectral variability (at least on timescales $\sim 10\ks$) can be
explained almost entirely by variations in the relative strength of
these two components, caused solely by changes in the normalisation of
the PLC (section~\ref{sect:rms}; see also Shih \et 2003 and Fabian \&
Vaughan 2003). An analysis of the flux-flux plots
(section~\ref{sect:ff}; see also Taylor \et 2003) and the difference
spectra (section~\ref{sect:diff}) shows the slope of the PLC remains
approximately constant with flux.

\item
The EPIC spectrum indicates there is resonance absorption by ionised
Fe at $\approx 6.7\keV$ (sections~\ref{sect:fe_abs} and
\ref{sect:line_core}). This was predicted based on the
presence of soft X-ray warm absorption (Matt 1994; Sako \et 2003) and
has been observed in at least one other high quality EPIC spectrum
(NGC~3783; Reeves \et 2003). This resonance absorption appeared to
vary between the two \xmm\ observations
(section~\ref{sect:big_diff}). 

\end{itemize}


\section{Discussion}
\label{sect:disco}

\subsection{Fitting the iron line in \mcg}

The broad iron line in \mcg\ has been unambiguously resolved using
\xmm/EPIC (W01; paper~I; this paper), confirming the previous results from
\asca\ (Tanaka \et 1995; Iwasawa 1996), \sax\ (Guainazzi \et 1999) and
\chandra\ (Lee \et 2002).   However, deriving reliable line profile
parameters is a considerable challenge even with the exceptionally
high quality \xmm$+$\sax\ data.  From the whole of the $320\ks$
observation the EPIC cameras collected about $6 \times 10^{3}$ counts from the
core of the line, peaking at $6.4\keV$, and about three times as many from
the broad red tail of the line emission.

The foremost problem is that there are no regions of the X-ray
spectrum unaffected by either the warm absorption (see
section~\ref{sect:wa_intro}) or the broad emission components. Thus it
is not possible to determine the underlying continuum without
simultaneously constraining the other spectral components.  Often
used, simple techniques for determining the continuum by e.g. fitting
a power-law over the $3-5\keV$ and $7-10\keV$ data (Nandra \et 1997),
will give a only crude first approximation of the continuum. In a
source such as \mcg, which possesses a strong warm absorber, the
absorption will cause the spectrum to curve even above $\sim 3\keV$
(see section~\ref{sect:wa_intro} and Fig.~\ref{fig:fluxed}).  In the
present paper this effect was dealt with by including an absorption
model derived from fits to the RGS data (section~\ref{sect:wa_intro})
and also allowing for additional absorption when fitting the EPIC data
(section~\ref{sect:simple}).

In addition to the distortion imposed on the continuum  the other
effect of warm absorption that has a significant impact on iron line
studies is resonant line absorption at $\approx 6.4-6.9 \keV$
(section~\ref{sect:fe_abs}).  The EPIC data indicate the presence of
such absorption; allowing for the Fe resonance absorption had a subtle
but significant effect on the best-fitting emission line models (see
section~\ref{sect:line_core}).  In particular, the line profile is
slightly broader and bluer after accounting for the line absorption.

Photoelectric absorption by Fe at energies $\sim 7.1-9.3 \keV$ could,
in principle, also be caused by the warm absorbing gas and further
confuse the continuum estimation (see e.g. Pounds \& Reeves 2002).
However, in the case of \mcg\ this is not a significant
effect.  The total opacity due to the Fe K edges, as predicted based
on fits to the soft X-ray RGS data, is negligible
(section~\ref{sect:fe_abs}), and  spectral analysis of the EPIC data
confirm this by ruling out the presence of deep Fe edges
(section~\ref{sect:fe_test}). The same is not true of all Seyfert 1s
galaxies. The columns of Fe ions derived using RGS data 
by Blustin \et (2002) for
NGC~3783, and by Sako \et (2001) for IRAS 13349+2438 predict strong K
edges with a total optical depth $\tau_{\rm max} \sim 0.05$ (see also
Reeves \et 2003).

Further complications include the presence of the reflection continuum
(section~\ref{sect:ref_intro}) and possible weak, narrow components to
the line emission (section~\ref{sect:core_intro}). The approach used
in the present paper was an attempt not to `unambiguously isolate' the
iron line but to model it after allowing for these complicating
factors.  In order to limit the range of models and free parameters,
constraints obtained from other instruments (such as the RGS and \sax\
spectra) were used where possible. The result of this spectral fitting
confirms the previous analyses in W01 and paper~I:  The best-fitting
emission model comprises a power-law continuum plus  strong reflection
from a weakly ionised, relativistic accretion disc
(section~\ref{sect:fit_with}).  This model requires the disc emission
to be centrally concentrated and extend in to $\sim 2\rg$. Attempts to
`explain away' the red wing of the line using enhanced absorption or
broken continua (section~\ref{sect:simple}) did not substantially
alter the main results. Additionally, alternative emission components
such as iron line blends did not fit the data
(section~\ref{sect:line_core}) while emission from soft X-ray lines or
blackbodies led to physically implausible results
(section~\ref{sect:fit_without}).

The only non-standard model that could fit the data with plausible
parameters was the partial covering model
(section~\ref{sect:fit_without}). However, as shown also by Reynolds
\et (2003),  this model is strongly at odds with the higher energy
data.  Thus, a partially covered continuum does not provide a
satisfactory explanation of the X-ray spectrum of \mcg. However, it is
possible that the source does contains some partially covered emission
in addition to the disc reflection.  Fitting the data with a
relativistic disc and also allowing for partial covering suggested the
possible impact of the partial covering is minimal
(section~\ref{sect:fit_without}). 

\subsection{Spectral variability of \mcg}

The spectral variability of the source can be explained in  terms of a
two-component model (section~\ref{sect:ff}; see also M$^{\rm c}$Hardy
\et 1998;  Shih \et 2002; Taylor \et 2003; Fabian \& Vaughan 2003).
This is the simplest model consistent with the flux-flux analysis.  

In this model the two spectral components are a power-law component
(PLC) and a reflection dominated component (RDC) which carries most of
the iron line flux. Both of these emission spectra are viewed through
the warm absorbing gas. Flux variations are caused primarily by
changes in the normalisation of the PLC, with the RDC remaining
relatively constant. Such a model explains many aspects of the
spectral variability including the linearity of the flux-flux plots
(section~\ref{sect:ff}), the rms spectrum (section~\ref{sect:rms}),
the correlation between $2-10\keV$ spectral slope and flux
(section~\ref{sect:fluxes}) and the lack of strong iron line
variations (Reynolds 2000; Vaughan \& Edelson 2001; paper~I).  The
relative constancy of the RDC causes variations above $\sim 2\keV$
(and particularly  around the iron line) to be suppressed, and also
variations below $\sim 1\keV$ are increasingly suppressed
(section~\ref{sect:rms}).  Fabian \& Vaughan (2003) previously applied
this model to direct fitting of the time-resolved EPIC spectra, where
it gave  a quite reasonable description of the time varying spectrum.

\begin{figure}
\rotatebox{270}{
\scalebox{0.35}{\includegraphics{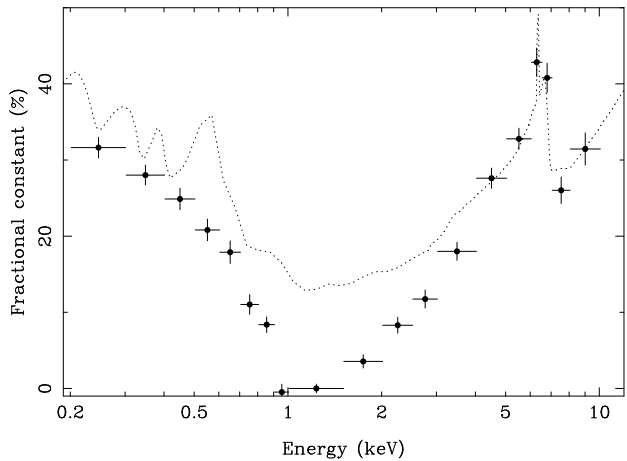}}}
\caption{
Fractional contribution to the spectrum of the constant component
deduced as in Fig.~\ref{fig:const}.  The dotted curve shows the
prediction based on the model derived from fitting the EPIC spectra in
the $3-10\keV$ range only. Clearly the model has qualitatively the
same shape, although the details do not match, particularly at lower
energies.  However, given that the model was derived without using the
spectral data below $3\keV$, and there is an intrinsic uncertainty in
the normalisation of the RDC (section~\ref{sect:ff}), the agreement is
rather interesting. 
}
\label{fig:rdc}
\end{figure}

The spectrum of the PLC was uncovered using the hi--low difference
spectrum (Fabian \& Vaughan; section~\ref{sect:fluxes}) and the
spectrum of the RDC was obtained  by isolating the constant offset of
the flux-flux plots (Taylor \et 2003; section~\ref{sect:ff}). This
latter technique revealed the spectrum of the constant RDC independent of any
spectral fitting, simply by analysing the light curves using flux-flux
plots. The result clearly revealed a spectral form strongly suggestive
of reflection, with a prominent iron line (as suggested by Fabian \&
Vaughan 2003). The RDC also shows a `soft
excess' below $\sim 1 \keV$ which could also be emission from the
reflecting disc  if it is weakly ionised. The model derived from
fitting the $3-10\keV$ EPIC spectra is in rough qualitative agreement
with the RDC spectrum obtained independently (see Fig.~\ref{fig:rdc}).
Thus the model derived from spectral fitting (section~\ref{sect:fit})
is in broad agreement with the model invoked to explain the spectral
variability (section~\ref{sect:spec_var}).

\subsubsection{Variability constraints on complex absorption models}

The warm absorption spectrum does not appear to vary substantially 
(specifically, the
optical depths of the absorption features remain approximately
constant).  For example, ratio plots of `high' and `low' spectral
(section~\ref{sect:fluxes}) show smooth profiles over the deepest
absorption features, as does the rms spectrum
(section~\ref{sect:rms}).  
This result is consistent with the studies of the deep
warm absorber in NGC~3783 using \chandra\ (Netzer \et 2003) and \xmm\
(Behar \et 2003) which showed the X-ray absorption remained unchanged
despite large changes in source luminosity.

Furthermore, the spectral variability constrained the possible effect
of warm or partial absorption on the iron line. Neither of the two
spectral components isolated through spectral variability analysis
resemble the absorbed part of the continuum in the partial covering model.
The depth of any line-of-sight Fe K edge was constrained by the
difference spectrum analysis (section~\ref{sect:fluxes}) to be 
$\tau \ls 0.02$ at $7.1\keV$ to $\tau \ls 0.07$ at $8.0\keV$. 
Fitting the partial covering model to the various difference spectra
gave constraints comparable to those based on the time-averaged
spectrum (section~\ref{sect:fit_without}). As discussed above the
effect of including such additional absorption (either warm or
partial) on the derived iron line parameters was negligible.

Alternative models for the spectral variability of \mcg\ in which
the variations are caused primarily by changes in the warm absorber (Nandra
\et 1990; Inoue \& Matsumoto 2003) are in stark contrast with the data.
This model predicts a power-law flux-flux relation (since the flux is
given by $f(E,t) = S(E) \exp(-\tau(E,t))$, where $S(E)$ is the underlying
the emission spectrum and $\tau(E,t)$ is the time-variable absorption
optical depth). Such a model would also predict strong features in the
rms spectrum corresponding to the deepest edges in the absorption
spectrum (paper~I). Both of these were ruled out.

\subsection{Outstanding issues with the disc model}

\subsubsection{Lack of iron line variations}

Although the model outlined above explains many aspects of the 
EPIC spectrum and its variability, there are some serious outstanding
problems. The most significant of these is how the RDC flux, including the
iron line, remains so steady in the presence of large changes in the
flux of the PLC. In standard accretion disc/reflection models the
reflection spectrum (including iron line) is fundamentally driven by
the PLC luminosity. Thus, since the RDC is thought to originate 
close to the SMBH (and the source of the PLC) the fluxes of the
two components should be tightly coupled.
The lack of correlations between the iron line and continuum is a
long-standing issue (Iwasawa \et 1996, 1999; Lee \et 2000; Reynolds
2000; Vaughan \& Edelson 2001). These \xmm\ observations demonstrate more clearly
than before that not just the line but the bulk of the reflected
emission (the RDC) shows little short-timescale variability.
The physics responsible for the reduced variability of the RDC is unclear. 

During the 2001 \xmm\ the total EPIC count rate from the iron line was
$\sim 0.1 $~ct s$^{-1}$, with $\sim 0.02 $~ct s$^{-1}$ coming from the
resolved, quasi-Gaussian core on the line and the rest coming from the
broad, red wing. As a result of this low count rate,
compounded by the modelling difficulties, it is difficult to  place
firm limits on weak and/or extremely rapid (faster than $\sim 10^4 \s$)
line variations.  Intriguingly, Reynolds \et (2003) claim that
variations in the line/RDC become apparent only when the source flux
is very low (see also Iwasawa \et 1996).

One possibility is the progressive ionisation of disc surface, as
discussed by e.g. Reynolds (2000), Nayakshin \& Kazanas (2002) and
Ballantyne \& Ross (2002). In this scenario an increase in the
ionising luminosity leads to little change in the flux of the line
because more of the Fe ions in the surface layers of the disc become
fully-stripped of electrons. Under some circumstances this model can
explain the lack of iron line variations (Reynolds 2000). However, the
spectral fitting analysis (section~\ref{sect:fit}) suggests the
emitting region of the disc is only weakly ionised. Time-resolved
spectral fitting of the \xmm\ data using an ionised disc model
(Ballantyne \et 2003) showed that ionisation effects could not
reproduce the constancy of the RDC. A second reflector that was
intrinsically constant was also required.  Thus, ionisation effects
alone are insufficient to explain the lack of iron line variations.

One intriguing alternative possibility is that gravitational light
bending, which may be strong if the disc extends to $\sim 2\rg$, is
(partially) responsible  (Fabian \& Vaughan 2003). See Miniutti \&
Fabian (2003) for a detailed discussion of this model.  Such a model
may naturally explain the enhanced reflection strength  ($R\gs 2$;
section~\ref{sect:ref_intro}) and iron line.

An interesting and related question is: how does the PLC retain its
spectral shape whilst undergoing factors of $\sim 5$ changes  in
luminosity? Both direct time-resolved spectral fitting (e.g. Fabian \&
Vaughan 2003) and flux-flux analyses 
(section~\ref{sect:ff}; Taylor \et 2003) show the slope of the PLC to stay remarkably
constant. Similar results have been seen in ultrasoft Seyfert galaxies
(e.g. Vaughan \et 2002) where the primary power-law slope stays
constant despite large amplitude luminosity variations.  This poses an
interesting challenge to models of the origin of the primary X-rays
which typically predict some luminosity-correlated variation in slope
(e.g. Zdziarski \et 2003). The observations are suggestive of some
kind of `Compton thermostat' (Haardt, Marachi \& Ghisellini 1994;
Pietrini \& Krolik 1995). Alternatively, if the apparent variations in
the PLC flux are due to geometrical effects near the SMBH, the PLC
might be expected to vary achromatically (Miniutti \et 2003).

\subsubsection{Inner edge of the disc}

The intricacies of modelling the complex spectrum of \mcg\ have
made it difficult to tap the potential of the iron line profile
for probing the region of strong gravity around the SMBH. 
An often asked question is: how far in towards the SMBH does the
disc extend? The answer to such a question may reveal the
spin of the SMBH (e.g. Stella 1990; Martocchia \et 2002).
The strength/shape of the red wing of the line is determined
by the emissivity of the innermost disc and so are the primary
diagnostics of the position of the disc's inner boundary. 
The problem is that the very broad, low-contrast tail to the
line is difficult to discern from the spectral curvature caused
by the warm absorber (see Fig.~\ref{fig:fluxed}). Thus it is not
possible to unambiguously  
determine the inner disc edge (unless the transmission of the warm
absorber is known to very high accuracy). 
For example, the method of Bromley, Miller \& Pariv (1998) works by
identifying the `minimum energy' of the red wing of the redshifted
line emission. However, no such energy can be identified without
detailed, simultaneous modelling of the absorption and the
reflection continuum.

This does not mean that all is lost.  The method of
fitting the reflection emission while simultaneously
fitting/constraining the absorbed continuum did require emission in to $\sim
2\rg$ (W01; paper~I; section~\ref{sect:fit_with}).  Importantly, this result
was robust to various possible effects that might bias the continuum
determination (sections~\ref{sect:simple} and
\ref{sect:fit_without}). This is strong but not conclusive evidence
that the SMBH is spinning. An alternative approach to fitting the red
wing of the line (perhaps isolating it through variability if the red
wing shows variations on very short timescales not yet probed)
would bolster the determination of its profile.  

The second and more fundamental objection is that some emission is
possible from within $6\rg$ even if 
the SMBH is not spinning (see Reynolds \& Begelman 1997; Young, Ross
\& Fabian 1998; Krolik \& Hawley 2002; Merloni \& Fabian 2003).
Further refinement of the theory combined with a further increase
in observational sensitivity should be able to settle this issue (see
Young \& Reynolds  2000; Yaqoob 2001).

\subsubsection{Why do only some Seyferts show broad lines?}

Recent, high-quality \chandra\ and \xmm\ observations of bright
Seyfert galaxies seem to have produced a mixed bag of iron lines.
Examples of strong, narrow lines have been found in NGC~3783 (Kaspi \et
2002; Reeves \et 2003), NGC 5548 (Yaqoob \et 2001; Pounds \et 2003a),
NGC 4151 (Schurch \et 2003), and NGC 3227 (Reeves 2003).  Examples of
highly broadened iron lines have been seen in  MCG--5-23-16 (Dewangan
\et 2003), NGC 3516 (Turner \et 2003c), IRAS 18325--5926 (Iwasawa \et
in prep.), Q~0056--363 (Porquet \& Reeves 2003), IRAS~13349+2438
(Longinotti \et 2003) and Mrk 766 (Mason \et 2003; Pounds \et 2003b).  
The line in \mcg\ is the broadest of the known broad lines.
There are some objects for which the line
profile is in dispute (e.g. Mrk 205 and Mrk 509, see Pounds \et 2001;
Reeves \et 2001; Page \et 2003c). Even more confusing is the possible
detection of rapidly  variable yet narrow iron lines in Mrk 841
(Petrucci \et 2002) and NGC 7314 (Yaqoob \et 2003).
Recent studies of GBHCs have also revealed relativistically broadened
iron lines in a few cases (Miller \et 2003; Miller \et 2002).

This raises two obvious, related questions. The first is: why do some
Seyferts show strong, broad lines while others show only narrow lines
(with any highly broadened component being weak or absent)? The second
question is: how can some Seyferts not have broad lines? 
The high luminosities of AGN in general argues that there must be
a large amount of mass flowing deep into the potential well of the
SMBH. Furthermore, the rapid X-ray variability argues
that the X-ray source is spatially compact and probably also located 
close to the SMBH. There should be a substantial amount of relatively
cool accreting matter in the region of strong gravity around the SMBH,
close to the primary X-ray source. These are the only two ingredients
thought necessary to produce a broad iron line (Fabian \et 1989;
Reynolds \& Nowak 2003). The absence of broad lines in many Seyfert 1s
is therefore quite puzzling and requires that either there simply is
not enough cool matter extending close to the SMBH to produce the line 
or that the line photons are somehow lost (absorbed/scattered) on their
way out of the inner regions. See also Reeves \et (2003) for a
discussion of this problem. 

It is important to note that there could be some observational biases
present. By their very nature highly broadened lines are difficult to
isolate (as discussed above). The broad line in \mcg\ is perhaps the
most clear example in part due to the strength of the reflection
spectrum. The high level of reflection ($R\gs2$) means that the
equivalent width of the broad line is rather high ($EW \sim 400
\eV$). Other Seyfert galaxies with more typical reflection strengths
($R \sim 1$) would be expected to have comparably weaker broad lines
($EW \sim 150 \eV$; George \& Fabian 1991). Even with the high
throughput of the EPIC cameras on-board \xmm, a typical length
observation ($\sim 40 \ks$) of a  bright Seyfert 1 might not clearly
reveal a redshifted line with $EW \sim 150 \eV$.  
This is especially true if the continuum is absorbed.
Thus, most
observations of Seyfert 1s are unlikely to show broad lines; the lack
of an obvious broad line in the spectrum does not rule out its
existence. 
These observations can constrain the strength of highly broadened lines
(e.g. upper limits on the equivalent width of a Laor profile) and so 
place useful limits on the line emission.
This will help address the question of whether broad
lines are a rare or common occurrence.

\subsection{Iron resonance absorption}

The fits to the iron line allowing for resonant absorption suggest the
presence of an additional 
absorber whose dominant effect is Fe~\textsc{xxv} He$\alpha$ absorption
at $\approx 6.7 \keV$.  
This absorption appears to have varied between the 2000 and 2001 \xmm\
observations (section~\ref{sect:big_diff}).
Such an absorption system can be modelled (using the {\tt
CLOUDY}  grids of Turner \et 2003a) with a high column ($\nh \gs
10^{22} \pscm$) and high ionisation ($\log(\xi) \approx 3$) warm
absorber.  This is similar to the results obtained for another bright
Seyfert 1,  NGC~3783 (Reeves \et 2003) except that this very high ionisation 
absorption does not significantly affect the low energy spectrum.
Similar features may have been observed in other Seyferts 
(NGC~3516, Nandra \et 1999; IRAS~13349-2438, Longinotti \et 2003).

Another approach is to estimate the column density using just the
equivalent width ($EW \sim 10 \eV$) of the absorption line and the
assumption that He-like Fe is the dominant ion in the absorbing
material. Using the oscillator  strength of the Fe~\textsc{xxv}
He$\alpha$ line from Nahar \& Pradhan (1999) the  column density can
be estimated assuming the line is unsaturated (linear part of the
`curve of growth'; Spitzer 1978).  This gives $N_{\rm Fe^{+24}} \sim
10^{17} \pscm$. If the line is saturated the column density will be higher.
The predicted depth of the corresponding Fe~\textsc{xxv} absorption
edge (at $8.83\keV$) is $\tau \sim 0.05$, consistent with the limits
obtained from the data (section~\ref{sect:fe_test}).  Assuming solar
abundances and that half all the Fe ions are in the He-like state this
column corresponds to $\nh \gs 4 \times 10^{21} \pscm$.  This compares
reasonably to the estimate derived from the {\tt CLOUDY} model, but
should really be considered only a lower limit on the total column
density since it does not account for other (lower  oscillator
strength) lines, possible line saturation or emission line  filling.

\subsection{The Future}

The long \xmm\ observation of \mcg\ has yielded a vast amount of
detailed information (on absorption and emission spectra, variability
timescales and delays and spectral changes). Only by making full use
of the long and simultaneous observations
(CCD and grating X-ray spectra)  possible with \xmm\ (and simultaneous
\sax\ observations) could all of these be achieved. Comparable
observations have been performed/planned for a small sample of other
Seyferts (for example: NGC 3783, Behar \et 2003; Reeves \et 2003)
which will shed light on the detailed X-ray properties of Seyferts in
general.  
Complimentary to these are the \chandra\ long-looks that 
have been particularly enlightening for warm absorber studies 
(e.g. Kaspi \et 2002).
Such observations also demonstrate the limits of present
instrumentation. For example, without higher resolution, high S/N spectroscopy
it is difficult to realise the diagnostic potential of the
$6.4-6.9\keV$ iron resonance absorption lines.

The presence of such a large column of highly ionised iron makes \mcg\
a good choice for high resolution Fe-band spectroscopy with the  XRS
on-board \astroe. This has the capability (with resolution $\Delta E
\sim 7 \eV$ and good throughput) to measure the depths and velocity
shifts of some of the individual resonance lines.  Indeed, the
combination of the high resolution XRS spectrum of the iron
absorption/emission, the high energy HXD spectrum which will constrain
the continuum and reflection, and the high throughput XIS spectrum
should provide the next major advance towards making full use of the
information contained in the X-ray spectra of Seyferts.

Aside from future missions, the present data indicate the most
interesting observations are often made when the source is very
faint. The spectral analysis presented above showed that when \mcg\ is
faintest, the iron line is strongest relative to the continuum, making it
a much higher contrast feature. The rare observations that 
show apparent changes in iron line properties tend to occur during
periods of unusually low flux (Iwasawa \et 1996; Reynolds \et 2003). 
Observations of NGC~4051 during its prolonged `low states' have
revealed other interesting phenomena, such as the warm absorbing gas
in emission (Uttley \et 2003). Thus it would seem that \xmm\ and
\chandra\ observations of bright Seyfert 1s, taken during 
periods of unusually low activity, may prove particularly revealing.


\section{Conclusions}
\label{sect:conc}

The long \xmm\ observation of \mcg\ has confirmed the existence of 
a highly broadened iron line. The main results of the present
investigation are as follows:

\begin{enumerate}

\item
It is not yet possible, given the present
state of knowledge of the detailed properties of the warm absorber, to
unambiguously dis-entangle the red wing of the line from the absorbed
continuum. 

\item
Nevertheless, the best-fitting model requires the
reflection/line emitting region to extend inwards to $\sim 2\rg$. 
This result was robust to various possible biases in the continuum
determination. 

\item
Alternative models for the spectral region around the iron line,
not allowing for the effects of strong gravity, gave unsatisfactory
results.

\item
The spectral variability of \mcg, including the correlation between 
$2-10\keV$ slope and flux and the reduced variability of the iron
line, has been explained using a two-component model. 

\item
The relative constancy of the reflection dominated component (RDC),
in the presence of large variations in the power-law component (PLC), 
is the primary cause of the spectral variations. The reasons for the
lack of RDC variations are unclear. A model invoking gravitational
light bending near the SMBH can qualitatively explain the 
suppressed variations, the relative strength of the RDC and the
small inner disc radius. 

\end{enumerate}


\section*{Acknowledgements}

Based on observations obtained with \xmm, an ESA science mission with
instruments and contributions directly funded by ESA Member States and
the USA (NASA). We thank an anonymous referee for a helpful report and
for suggesting the PCA analysis. 
We thank Adrian Turner for help with the warm absorber
models, David Ballantyne and Randy Ross for help with the reflection
models, Giovanni Miniutti for help understanding the GR effects, and
Kazushi Iwasawa for many illuminating discussions.  We also thank
Steve Sembay for useful discussions about the EPIC calibration and
Phil Uttley for advice about spectral variability analysis.  SV 
thanks PPARC for support.  ACF thanks the Royal Society for support.
This research has made use of NASA's Astrophysics Data System and the
NASA/IPAC Extragalactic Database (NED) which is operated by the Jet
Propulsion Laboratory, California Institute of Technology.


\appendix

\section{Analysis of EPIC calibration data}
\label{app:cal}

This section describes an analysis of EPIC data taken from
observations of \3c\ ($z=0.158$). 
This source is bright,  point-like
(compared to the EPIC PSF) and shows a relatively  simple energy spectrum
in the EPIC bandpass (i.e. approximately power-law with no sharp
spectral features), making it useful for assessing the
accuracy of the EPIC spectral calibration.

Five observations of \3c\ were analysed. These observations were 
taken primarily for performance verification and calibration purposes
and the observation details are given in Table~\ref{tab:cal}. 
During all five observations the source was positioned on-axis
and all three EPIC cameras used small window mode and medium filter
(i.e. the same instrumental set-up as the \mcg\ observations). 
The only exception is the first half of the rev. $277$ observation
during which the MOS cameras used the thin filter (data from this part
of the observation were ignored). This later observation was used by
Molendi \& Sembay (2003) to assess the EPIC calibration.

\begin{table*}
\begin{center}
\caption{Details of observations used for calibration purposes.}
\begin{tabular}{lllll}                
\hline
          &      & Observation    & Observing time & Duration    \\
Source    &Revolution &    ID          &  (UT)         &       (s)     \\
\hline
\3c\      & 094  & 0126700301     & 2000 Jun 13 23:39:53--2000 Jun 14 20:05:49 & 73556      \\
\3c\      & 095  & 0126700601     & 2000 Jun 15 12:58:18--21:35:30             & 31032      \\
\3c\      & 095  & 0126700701     & 2000 Jun 15 23:32:02--2000 Jun 16 09:37:48 & 36346      \\
\3c\      & 096  & 0126700801     & 2000 Jun 17 23:24:14--2000 Jun 18 19:50:15 & 73561      \\
\3c\      & 277  & 0136550101     & 2001 Jun 13 07:14:26--2001 Jun 14 08:10:31 & 89765      \\
\hline
\label{tab:cal}
\end{tabular}
\end{center}
\end{table*}

\begin{figure}
\rotatebox{270}{
\resizebox{!}{\columnwidth}{\includegraphics{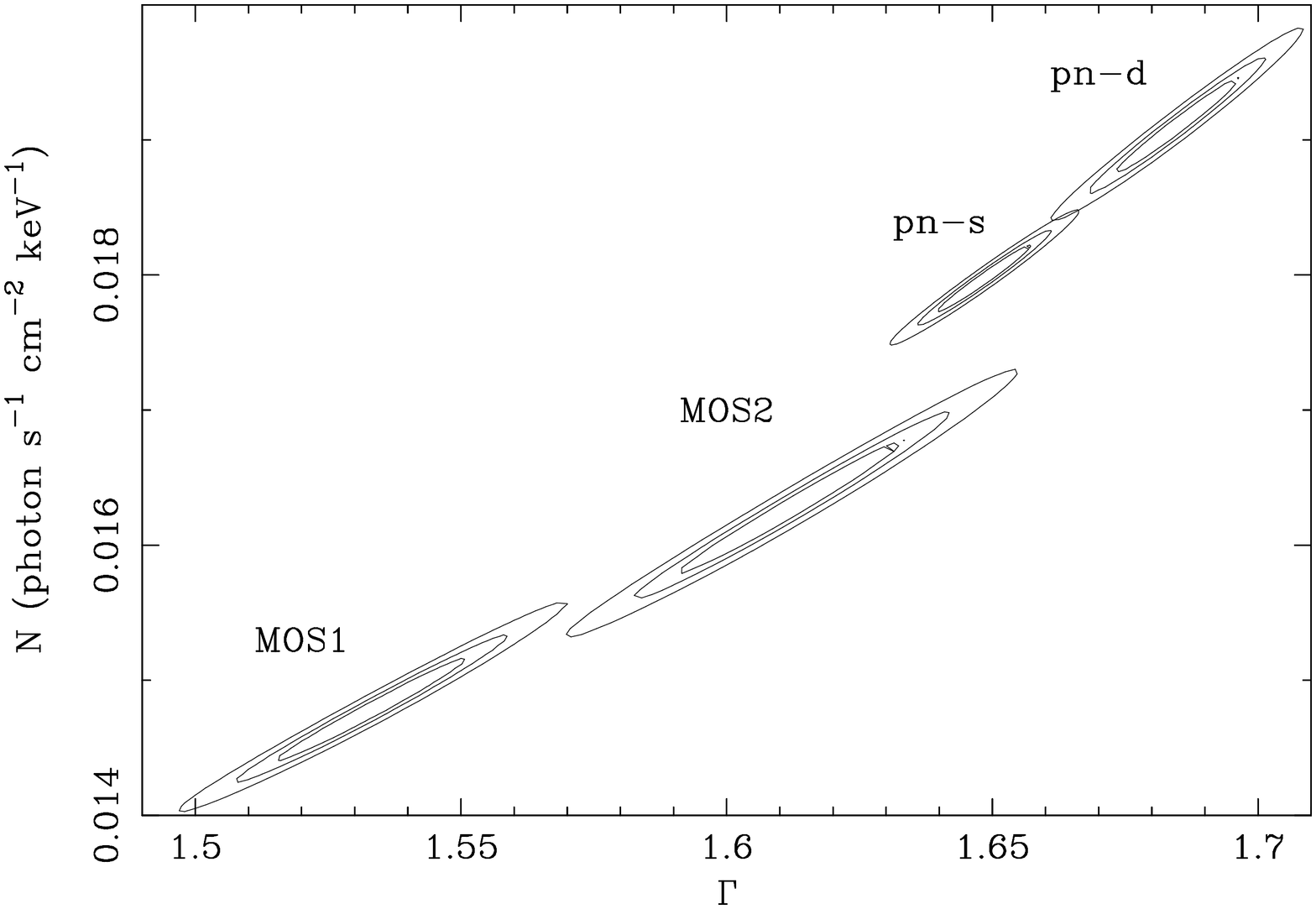}}}
\caption{
Confidence contours from fitting the \3c\ EPIC
spectra (from revs. $0094-0096$) over the $3 - 10\keV$ range 
with a power-law model.
}
\label{fig:3c273_cont1}
\end{figure}

\begin{figure}
\rotatebox{270}{
\resizebox{!}{\columnwidth}{\includegraphics{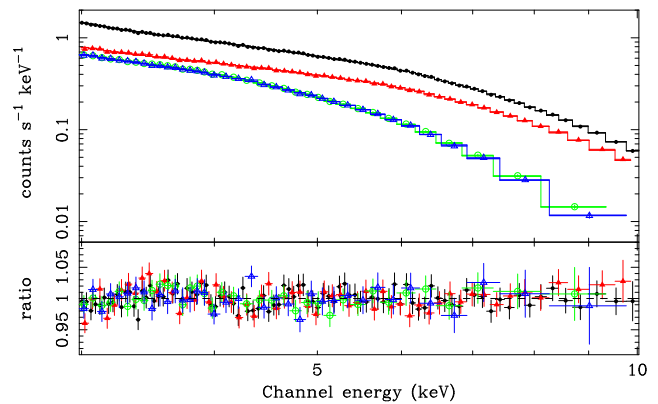}}}
\caption{
$3 - 10\keV$ EPIC spectra of \3c from revs. $094 - 096$.
The four sets of data are for pn single (filled circles) and double
events (filled triangles), MOS1 (hollow circles) and MOS2 (hollow triangles)
The spectra were fitted with a model comprising a
power-law and a Gaussian emission line.
}
\label{fig:3c273_spec1}
\end{figure}

Source and background spectra were extracted exactly as for the \mcg\
observation.  Only single pixel event were used from the MOS to
reduce the effect of pile-up in this bright source (see Molendi \&
Sembay 2003).  The grouped spectra were then fitted in the $3 -
10\keV$ range with a simple power-law modified by Galactic absorption
($\nh=1.79 \times 10^{20}\pscm$; Dickey \& Lockman 1990).
The power-law indices and normalisations were left independent for
each camera.  In all five observations the spectra from each camera
showed weak but positive residuals at $5 - 6\keV$, suggesting the
presence of iron line emission. Previous studies of \3c\ (e.g. Turner
\et 1990; Grandi \et 1997; Reeves 2003) also suggested the presence of
weak iron emission. A Gaussian was therefore included in the model at
$6.4\keV$ (in the source frame) to account for iron emission (with
equivalent width $EW \sim 20\eV$).  This simple model provided an
acceptable  fit to all five sets of spectra
(i.e. $\chi_{\nu}^{2}<1.0$) and showed that the four sets of spectra
taken during rev. $094 - 096$ were very similar (i.e. the source
showed no spectral variability during this period).  Therefore the
data from the four observations spanning these revolutions were
combined to produce a high signal-to-noise spectrum for each EPIC
camera.


Fig.~\ref{fig:3c273_cont1} shows a comparison of the photon indices
and normalisations for the three EPIC spectra from the rev. $094 -
096$ data.  Similar results were obtained from the rev. $277$ data.
Clearly the spectral indices are discrepant, with the MOS1 spectral
index significantly lower than MOS2, which in turn is lower than the
pn.   The flatter index derived from MOS1 is a known calibration
problem (Molendi \& Sembay 2003), and the other three spectra are
within $\Delta \Gamma \approx 0.07$ of each other.  As there is likely
to be some spectral distortion due to pile-up  in the MOS data, even
in the single pixel events, the differences in slope between the MOS
cameras and the pn might be expected to  be slightly smaller for a
fainter (i.e. less piled-up) source such as \mcg.
Fig.~\ref{fig:3c273_spec1} shows the data and residuals from the
power-law plus Gaussian fit. Clearly this simple model reproduced the
EPIC spectra to better than $\pm 5$ per cent over the $3-10\keV$
range.  Excepting these systematic differences in spectral slope
between the three cameras, the EPIC spectral calibration appears good
to $\sim 5$ per cent in the $3 - 10\keV$ band
(Fig.~\ref{fig:3c273_spec1}).
The only noticeable sharp residual in the $3-5.5\keV$ range occurs 
at $\approx 4\keV$ in the observed frame. A shallow absorption
edge fits this feature ($E=4.14_{-0.17}^{+0.08}\keV, ~ \tau =
0.023_{-0.005}^{+0.011}$). This is possibly an instrumental feature
caused by Ca deposits on the mirror surfaces, but is only present as a
$\sim 3$ per cent effect.

Extrapolating the above model down to lower energies reveals and
upturn in the data, i.e. \3c\ has a `soft excess'. The $0.5 - 3\keV$
spectrum can be fitted with a broken power-law or two blackbody
components to account for the curved
continuum. The parameters were allowed to vary independently for each
camera and again the spectral slopes were significantly different. The fit
was poor but the residuals are generally confined to the $\pm 4$ per
cent level. The most noticeable exceptions are near the instrumental
edges (O K-edge at 0$.538\keV$, Si K-edge at $1.846\keV$ and Au M-edge
at $2.21\keV$), particularly evident in the high S/N pn spectra.

\bsp
\label{lastpage}
\end{document}